\documentclass[journal]{IEEEtran}

\usepackage{graphicx}
\usepackage{url}
\usepackage{amsmath,amssymb,exscale}
\usepackage[compress]{cite}

\usepackage{color}
\usepackage{graphicx}
\usepackage{algorithm}
\usepackage{algorithmic}
\usepackage[center]{caption}
\usepackage{verbatim}
\usepackage[bookmarks=false,hidelinks]{hyperref}
\usepackage{balance}

\usepackage{booktabs}
\usepackage{array}
\usepackage{multirow}
\usepackage{resizegather} 
\usepackage{bbm}
\usepackage{subcaption}
\usepackage{lineno}

\definecolor{reviewold}{rgb}{0.65,0.08,0.08}

\definecolor{reviewcontext}{rgb}{0.38,0.38,0.38}
\makeatletter
\newcommand{\writingreviewcontext}{\@fileswfalse\color{reviewcontext}}
\makeatother
\usepackage{amsthm}
\newtheoremstyle{italic-head-normal-body}
  {\topsep}{\topsep}     
  {\normalfont}           
  {}                      
  {\itshape}              
  {}                     
  {0.5em}                 
  {\thmname{#1}~\thmnumber{#2}: \thmnote{(#3)}} 

\theoremstyle{italic-head-normal-body}
\newtheorem{theorem}{Theorem}
\begin{document}

\captionsetup[figure]{name={Fig.},labelsep=period,singlelinecheck=off} 
\graphicspath{{Images/}}
\renewcommand{\algorithmicrequire}{\bf{Input:}}
\renewcommand{\algorithmicensure}{\bf{Output:}}

\title{\resizebox{\textwidth}{!}{Semantic Sensing: Toward a Task-Oriented Paradigm}}

\author{Xiaoqi Zhang,~\IEEEmembership{Student Member,~IEEE,}
        J. Andrew Zhang,~\IEEEmembership{Senior Member,~IEEE,}
        Chang Liu,~\IEEEmembership{Member,~IEEE,}
        Weijie Yuan,~\IEEEmembership{Member,~IEEE,}
        Geoffrey Ye Li,~\IEEEmembership{Fellow,~IEEE,}
        and Moeness G. Amin, ~\IEEEmembership{Life Fellow, IEEE}

\thanks{X. Zhang and J. A. Zhang are with the School of Electrical and Data Engineering, University of Technology Sydney, Sydney, NSW 2007, Australia (e-mail: Xiaoqi.Zhang@student.uts.edu.au; andrew.zhang@uts.edu.au).}
\thanks{C. Liu is with the Department of Computer Science and Information Technology, La Trobe University, Melbourne, Victoria 3086, Australia (e-mail: C.Liu6@latrobe.edu.au).}
\thanks{W. Yuan is with the School of Automation and Intelligent Manufacturing, Southern University of Science and Technology, Shenzhen 518055, China (e-mail: yuanwj@sustech.edu.cn).}
\thanks{G. Y. Li is with the Department of Electrical and Electronic Engineering, Imperial College London, London SW7 2AZ, U.K. (e-mail: geoffrey.li@imperial.ac.uk).}
\thanks{Moeness G. Amin is with the Center for Advanced Communications, Villanova University, Villanova, PA 19085 USA (e-mail: moeness.amin@villanova.edu).}
\thanks{(Corresponding author: J. Andrew Zhang.)} 
}

\maketitle
% % BEGIN DELETION REVIEW LEGEND
% {\color{reviewold}\footnotesize\noindent
% \textbf{Author review copy.} Remaining Writing comparison blocks compare the wording before the
% Writing revisions (2bcd05e) with the current text immediately below each block.
% Block P63 retains the previous pilot-budget discussion.
% Within these blocks, red bold text marks deleted or replaced wording;
% gray text retains context. Blue text in the current manuscript marks revisions.
% The earlier reviewer block retains its original baseline (13671b8).
% Item labels R1-/R2- identify Reviewer 1/Reviewer 2 within Round 1;
% M017/M018 identify internal items.
% Response page locators refer to the revised text in
% \texttt{main\_R1\_highlight.pdf}, which excludes these red blocks.
% \par}
% % END DELETION REVIEW LEGEND

\begin{abstract}
Sensing and communication are fundamental enablers of next-generation networks. While communication technologies have advanced significantly, the potential of sensing beyond conventional parameter estimation remains far from fully explored. To address these limitations, we propose semantic sensing (SemS), a framework that extends task-oriented sensing through explicit modeling and recovery of task-relevant environmental information. Specifically, we mathematically formulate how transmit waveforms and receiver processing jointly support the recovery of task-relevant semantics, thereby establishing SemS as a semantics-oriented transceiver design. Within this architecture, we formulate an information-theoretic objective that guides the sensing pipeline to maximize task-relevant information extraction under resource constraints. To solve this optimization problem, we develop a deep learning (DL)-based framework that jointly designs transmit waveform parameters and receiver representations. The framework is implemented in an orthogonal frequency division multiplexing (OFDM) system, featuring a semantic encoder that employs differentiable pilot selection to select informative time--frequency resources. At the receiver, we design a shared semantic decoder that estimates three semantic attributes, namely object category, proximity, and radial motion, to support target classification and approach warning through predefined task mappings. We also consider continuous target range estimation as a special case in which the semantic state and task output coincide. Numerical results demonstrate that the proposed semantic pilot design achieves superior classification accuracy and approach-warning Macro-F1 compared to reconstruction-based baselines, particularly under constrained resource budgets.
\end{abstract}


\IEEEpeerreviewmaketitle

\section{Introduction}
As a key enabler for beyond fifth-generation (B5G) networks, integrated sensing and communication (ISAC) is expected to support both ubiquitous connectivity and advanced sensing capabilities for a wide range of emerging applications, such as autonomous driving, industrial robotics, and unmanned platforms \cite{liu2022integrated, zhang2021enabling, amin2026special}. However, conventional ISAC primarily prioritizes signal-level fidelity through accurate recovery of transmitted symbols and physical measurements. Although this supports high-fidelity reconstruction, intelligent applications require semantic understanding for tasks, such as object recognition and scene interpretation \cite{schumann2018semantic}. Consequently, these applications motivate a shift toward semantic interpretation. \par
This transition is particularly evident in the communication domain, where semantic communication (SemCom) has begun to integrate high-level semantics into wireless systems, shifting the design objective from bit transmission to meaning exchange. In this context, semantics is defined as task-oriented information abstracted from raw observations that directly supports decisions by intelligent systems \cite{qin2021semantic}. Related advances in deep learning (DL)-based joint source-channel coding (JSCC) have improved image reconstruction over noisy communication channels compared with separation-based schemes, particularly at low signal-to-noise ratios (SNRs) \cite{bourtsoulatze2019deep}. \par
% Furthermore, recent studies highlight the potential of semantic-aware resource allocation to enhance the spectral efficiency of such systems \cite{gunduz2022beyond}. \par
While extending semantics to sensing remains an active research direction, cognitive radar provides a foundation through a biologically inspired perception-action cycle that closes the feedback loop between the transmitter and environment \cite{haykin2006cognitive}. Building upon this principle, adaptive waveform design and array reconfiguration use environmental knowledge to guide sensing \cite{martone2021view, wang2023enhanced}. For instance, waveform optimization techniques have been employed to minimize the mean-squared error (MSE) of target parameter estimation \cite{yang2007mimo} or to maximize the signal-to-interference-plus-noise ratio (SINR) in the presence of signal-dependent clutter \cite{naghsh2014radar}. More recently, predictive beamforming approaches using spatio-temporal channel state information (CSI) have been shown to improve sensing accuracy in terms of the Cramér–Rao bound (CRB) \cite{zhang2024predictive}. Cognitive radar also supports task-driven acquisition, including adaptive waveform design for target recognition \cite{wei2010extended}. End-to-end radar learning has further combined supervised detector training with reinforcement learning of transmit waveforms \cite{jiang2019end}. Within this cognitive sensing perspective, semantic learning focuses on the higher-level representation and recovery of task-relevant environmental information.\par

% semantic in sensing system

Driven by the demand for high-level semantics in intelligent applications, radar sensing has gradually shifted its focus from physical parameter estimation to semantic tasks \cite{shi2022human,held2018radar,schumann2018semantic}, including target classification and behavioral analysis. To facilitate these objectives, DL models are usually deployed at the receiver side to infer task-oriented information directly from the received signals. Specifically, one-shot learning schemes have been developed for semantic-level activity recognition in WiFi-based sensing, enabling accurate classification even in previously unseen scenarios \cite{shi2022human}. Similarly, micro-Doppler signatures from frequency-modulated continuous-wave (FMCW) radar have been extensively employed to perform human daily activity recognition as well as gesture and sign language recognition, demonstrating that semantic information can be extracted by using motion patterns rather than relying on physical parameters alone \cite{gurbuz2020american, gurbuz2019radar,amin2019hand}. Radar and synchronized motion-capture measurements have also been used to analyze pedestrian micro-Doppler signatures and limb-motion features in road-crossing scenarios \cite{held2018radar}. Furthermore, DL algorithms have been utilized in autonomous driving to segment high-dimensional radar data into distinct object classes for downstream decision-making \cite{schumann2018semantic}. These advancements align with ``Radio Vision,'' where the primary objective shifts towards extracting application-oriented semantic information \cite{zhang2020perceptive}. This shift toward semantic interpretation raises a fundamental design question. \textit{How can we jointly design the waveform and receiver processing to extract the information most relevant to the sensing task?}\par

% Semantic ISAC
While the emerging ISAC framework theoretically offers the degrees of freedom to address the underlying issue, current studies on integrating semantics into ISAC have predominantly adopted a communication-centric perspective. In these paradigms, the integration is typically realized as ``semantic communication for sensing,'' where the primary objective is to employ semantic coding techniques for efficient compression and transmission of sensed data, such as radar point clouds or images, to a receiver \cite{lu2024semantic,cao2025task}. Within this model, the sensing module functions primarily as a source of sensed data for communication, with optimization targets restricted to transmission efficiency, reconstruction fidelity, or receiver-side adaptive sampling \cite{zhang2022semantic}. This communication-centric approach focuses on transmitting acquired information rather than actively probing the environment. Consequently, the transmit waveform is not adapted to downstream semantic tasks. \par

% Semantic Sensing Definition
To bridge this gap, we propose semantic sensing (SemS), a framework that extends task-oriented sensing by incorporating semantic representation and recovery into joint waveform and receiver design. Inspired by environmental semantic characterization \cite{zhang2025channel}, we view the sensing channel as a source of information about target attributes, behaviors, and events reflected in multipath propagation. We establish the SemS transceiver architecture and formulate an information-theoretic objective for semantic recovery under resource constraints. To realize this objective, we develop a DL-based optimization method that jointly learns the waveform and receiver parameters. We implement and evaluate this method in an orthogonal frequency division multiplexing (OFDM) system using learned pilot selection and a shared semantic decoder for target classification and approach warning. The main contributions of this paper are summarized as follows.
\begin{itemize}
    \item We propose SemS as an extension of task-oriented sensing that explicitly models the acquisition and recovery of semantic information for downstream inference. Specifically, we formulate the relationship between task-relevant environmental semantics, waveform-dependent physical observations, and semantic recovery to support joint waveform and receiver design. Building on this formulation, we establish the SemS architecture and define its core components. Moreover, we formulate an information-theoretic criterion for jointly optimizing the sensing pipeline toward the semantic objective under practical resource constraints.
    \item We develop a cross-layer, end-to-end optimization methodology that jointly optimizes transmit resource allocation and receiver interpretation. By bridging the physical and semantic domains, this approach bypasses the full-channel reconstruction stage used in reconstruction-oriented sensing pipelines, directly linking physical waveform parameters to the semantic-recovery loss. The formulation supports efficient gradient-based optimization through the differentiable channel model while retaining a modular structure that accommodates diverse waveform constraints and receiver architectures.
    \item We realize the proposed framework in an OFDM system by optimizing pilot locations on the time--frequency grid subject to a constraint on the number of pilots. A differentiable approximation to discrete pilot selection enables joint optimization of the pilot pattern and semantic decoder using cross-entropy losses for semantic-attribute recovery. Predefined task mappings use the predicted attribute probabilities to produce target-classification and approach-warning outputs. The resulting pilot pattern and decoder parameters are reused for both tasks without task-specific retraining. Simulation results demonstrate that, for the same number of pilots, the proposed design achieves higher target-classification accuracy and approach-warning Macro-F1 than reconstruction-based pilot design, with the classification advantage retained under overlapping radial-motion distributions. For continuous target range estimation, where the semantic state and task output coincide, joint pilot and receiver optimization achieves lower mean absolute error than the compared baselines in clutter.
% Numerical results demonstrate that the proposed semantic pilot design yields superior task accuracy and estimation precision compared to reconstruction-based benchmarks, particularly in resource-constrained regimes.
\end{itemize} \par
% T1--T3 approved 2026-09-19.
% S-01/S-02 and physical-channel-only terminology approved for implementation 2026-09-20.
% M-01 complete-path input and S-03 replacement example approved 2026-09-20.
% Clean manuscript approved 2026-09-22; revision colors remain in main_R1_highlight.tex.
The remainder of this paper is organized as follows. Section~\ref{sec:physical_model_semantics} presents the physical sensing model and defines task-relevant semantics. Section~\ref{sec:semantic_sensing_framework} introduces the SemS architecture and transceiver co-design, characterizes semantic recovery, and formulates the information-theoretic design criterion. Subsequently, Section \uppercase\expandafter{\romannumeral4} details the end-to-end optimization methodology. Section \uppercase\expandafter{\romannumeral5} implements the framework through OFDM-based pilot design for target classification and \mbox{approach} warning. Section \uppercase\expandafter{\romannumeral6} presents the simulation results, and Section \uppercase\expandafter{\romannumeral7} concludes the paper. \par

\textit{Notation:} Throughout this paper, scalars are written in plain letters, vectors in bold lowercase, and matrices in bold uppercase. \(\mathbb{R}^{M \times N}\) and \(\mathbb{C}^{M \times N}\) denote the sets of real-valued and complex-valued matrices of dimension \(M \times N\), respectively. For a complex number, \(\Re(\cdot)\) and \(\Im(\cdot)\) indicate its real and imaginary parts. A real Gaussian distribution with mean vector \(\boldsymbol{\mu}\) and covariance matrix \(\boldsymbol{\Sigma}\) is denoted by \(\mathcal{N}(\boldsymbol{\mu}, \boldsymbol{\Sigma})\), while a circularly symmetric complex Gaussian distribution is written as \(\mathcal{CN}(\boldsymbol{\mu}, \boldsymbol{\Sigma})\). $\odot$ denotes the Hadamard product. Operators \((\cdot)^{T}\), \((\cdot)^{H}\), and \((\cdot)^{*}\) denote transpose, Hermitian transpose, and complex conjugation, respectively.

\begin{figure}[tb]
    \centering
    \includegraphics[width=0.85\linewidth]{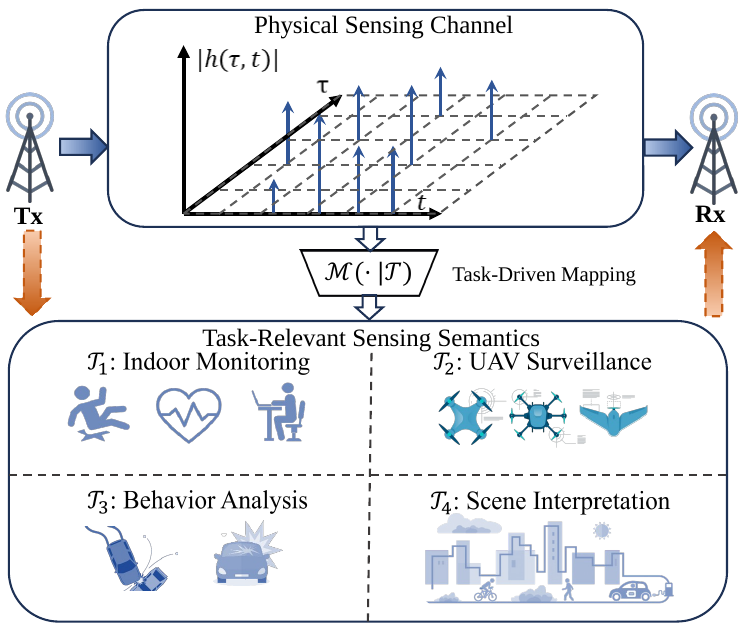}
    \caption{Illustration of task-conditioned mappings from the physical sensing environment to task-relevant semantics.}
    \label{fig:sem_channel}
\end{figure}

\section{System Model}
\label{sec:physical_model_semantics}
% BEGIN R1-4/R1-5; M017/M018 model dependencies
We consider a digital baseband system unless stated otherwise. In this section, we describe the physical sensing environment through its observation model and path parameters. Task-relevant semantics are then defined as an abstraction of this physical state that supports downstream inference.
\subsection{Physical Sensing Model}
\label{sec:physical_sensing_model}
We consider a general sensing system equipped with $N_t$ transmit and $N_r$ receive antennas in a multiple-input multiple-output (MIMO) configuration. Physical propagation is described by a discrete-time complex-baseband linear time-varying (LTV) model. Let $x_p[k]$ and $y_q[k]$ denote the transmitted and received samples at antennas $p$ and $q$, respectively, where $p=1,\ldots,N_t$ and $q=1,\ldots,N_r$. The received signal at antenna $q$ is
{
\begin{equation}
y_q[k]=\sum_{p=1}^{N_t}\sum_{m\in\mathbb Z}h_{q,p}[k,m]x_p[k-m]+n_q[k],
\label{eq:ltv_channel}
\end{equation}
}%
where $h_{q,p}[k,m]$ denotes delay tap $m$ of the sampled channel impulse response (CIR) from transmit antenna $p$ to receive antenna $q$ at sample $k$, and $n_q[k]$ denotes additive receiver noise \cite{tse2005fundamentals}. {\color{black}
Over each processing interval, we approximate path gains, delays, angles, and Doppler shifts as constant \cite{andrew2021overview}. For far-field propagation with negligible beam squint and negligible Doppler-induced phase variation over the effective pulse duration, the CIR between antennas $p$ and $q$ is
\par
{\small
\begin{equation}
h_{q,p}[k,m]=\sum_{l=1}^{L}\alpha_l a_{r,q}(\theta_l)a_{t,p}(\vartheta_l)e^{j2\pi\nu_l kT_s}g\!\left(m-\frac{\tau_l}{T_s}\right),
\label{eq:discrete_cir}
\end{equation}
}%
where $T_s$ is the sampling interval and $L$ is the number of modeled paths. For path $l$, $\alpha_l$, $\tau_l$, $\nu_l$, $\theta_l$, and $\vartheta_l$ denote the complex gain, delay, Doppler shift, angle of arrival (AoA), and angle of departure (AoD), respectively. Factors $a_{r,q}(\theta_l)$ and $a_{t,p}(\vartheta_l)$ are entries of the receive and transmit array-response vectors, using $\mathbf a_r(\theta_l)\mathbf a_t^T(\vartheta_l)$ for the spatial response. The function $g(u)$ is the effective pulse response from transmit pulse shaping and receive filtering, with delay measured in sampling intervals.

For given array responses and pulse shaping, the physical state of path $l$ is $\mathbf p_l=[\tau_l,\nu_l,\theta_l,\vartheta_l,\alpha_l]^T$. The path set, $\mathcal P_{phy}=\{\mathbf p_l\}_{l=1}^{L}$, contains all modeled target and clutter paths over the observation interval.\par
}

\subsection{Task-Relevant Sensing Semantics}
\label{sec:task_relevant_semantics}
To bridge the gap between raw physical observations and high-level semantics, we formulate the semantic representation $\mathcal S_{\mathrm{sem}}^{\mathcal T}$ as a task-oriented abstraction of the physical environment through
\begin{equation}
\mathcal S_{\mathrm{sem}}^{\mathcal T}=\mathcal M(\mathcal P_{phy}\mid\mathcal T),
\label{eq:semantic_projection}
\end{equation}
% BEGIN R2-2 domain clarification
where $\mathcal M$ is a generally nonlinear, task-conditioned mapping on the complete path set, $\mathcal P_{phy}$, with $L\ge1$.
% END R2-2 domain clarification
% BEGIN R2-4 task-role clarification
Here, $\mathcal T$ denotes a sensing task specification comprising a single task or a predefined set of tasks, and $\mathcal S_{\mathrm{sem}}^{\mathcal T}$ denotes the true semantic state. For an individual task, the true task output for a given physical realization is $T_o=\mathcal M_T(\mathcal S_{\mathrm{sem}}^{\mathcal T})$, where $\mathcal M_T$ denotes the task-specific inference function. For example, in the indoor-monitoring application illustrated in Fig.~\ref{fig:sem_channel}, the semantic state can represent human activity states such as sitting, walking, or falling. In this example, $\mathcal M_T$ maps sitting and walking to a safe category and falling to a danger category. In addition,
Table~\ref{tab:semantic_mapping_classes} summarizes representative forms of the physical-to-semantic mapping and a learned realization. The first five forms can also be implemented by neural networks or foundation models adapted to the sensing data and task. Here, $f$ denotes a task-specific function in each row, $\rho$ maps summed path features to semantics, and $\psi$ parameterizes $\mathcal M_\psi$. Index $t$ labels observation intervals, and $W$ denotes the window length in intervals.
\par
% END R2-4 task-role clarification

% BEGIN S-03 region-occupancy example; replaces material recognition
Following recent characterizations of environmental semantics in wireless propagation \cite{zhang2025channel,zhang2024ai}, we consider status, behavior, and event semantics \cite{zhang2025channel}. For example, region occupancy provides status semantics, similar to the region-based formulation in \cite{mojahedian2026task}. Consider $\mathcal T=$ intrusion detection in a two-dimensional monostatic scene with direct point-target reflections. Let $\{\mathcal R_j\}_{j=1}^{J}$ denote the monitored regions and $\mathcal I_{\mathcal T}\subseteq\{1,\ldots,L\}$ index the associated direct target reflections within $\mathcal P_{phy}$. With the sensor at the origin, path $l\in\mathcal I_{\mathcal T}$ has position $\mathbf r_l=(c\tau_l/2)[\cos\theta_l,\sin\theta_l]^T$, where $c$ is the speed of light. The occupancy of region $j$ is
\begin{equation}
o_j=\mathbbm{1}\!\left\{\exists\,l\in\mathcal I_{\mathcal T}:\mathbf r_l\in\mathcal R_j\right\},
\label{eq:region_occupancy}
\end{equation}
where $\mathbbm{1}\{\cdot\}$ is the indicator function. The semantic state is
\begin{equation}
\mathcal S_{\mathrm{sem}}^{\mathcal T}=\mathcal M(\mathcal P_{phy}\mid\mathcal T)=[o_1,\ldots,o_J]^T,
\label{eq:occupancy_semantic_state}
\end{equation}
and intrusion in regions indexed by $\mathcal J_{\mathrm r}\subseteq\{1,\ldots,J\}$ is determined by
\begin{equation}
T_o=\mathcal M_T(\mathcal S_{\mathrm{sem}}^{\mathcal T})=\mathbbm{1}\{\textstyle\sum_{j\in\mathcal J_{\mathrm r}}o_j>0\}.
\label{eq:intrusion_decision}
\end{equation}
For example, $\mathcal S_{\mathrm{sem}}^{\mathcal T}=[1,0,1]^T$ indicates that regions 1 and 3 are occupied and region 2 is unoccupied. For $\mathcal J_{\mathrm r}=\{3\}$, region 3 is the region of interest, and $T_o=1$ indicates an intrusion in this region.
% END S-03 region-occupancy example

% END E-02 physical-to-semantic mapping examples

\begin{table}[!t]
\centering

\caption{Representative Physical-to-Semantic Mappings}
\label{tab:semantic_mapping_classes}
\footnotesize
\setlength{\tabcolsep}{3pt}
\renewcommand{\arraystretch}{1.06}
\begin{tabular}{@{}>{\raggedright\arraybackslash}p{.28\columnwidth}>{\raggedright\arraybackslash}p{\dimexpr.72\columnwidth-2\tabcolsep\relax}@{}}
\toprule
\textbf{Mapping} & \textbf{Expression and example} \\
\midrule
Parameter selection &
\(\mathcal S_{\mathrm{sem}}^{\mathcal T}=\{(\tau_l,\nu_l)\}_{l\in\mathcal I_{\mathcal T}}\)\par
Retain target delays and Dopplers for unmanned aerial vehicle (UAV) tracking. \\
\specialrule{0.3pt}{2pt}{2pt}
Nonlinear mapping &
\(\displaystyle \mathcal S_{\mathrm{sem}}^{\mathcal T}=\frac{c\tau_l}{2}[\cos\theta_l,\sin\theta_l]^T\)\par
Convert the delay and AoA of a direct monostatic return to position for UAV localization. \\
\specialrule{0.3pt}{2pt}{2pt}
Cross-path aggregation &
\(\mathcal S_{\mathrm{sem}}^{\mathcal T}=\rho\!\left(\sum_{l\in\mathcal I_{\mathcal T}}f(\mathbf p_l)\right)\)\par
Combine target-path features to identify occupied indoor regions. \\
\specialrule{0.3pt}{2pt}{2pt}
Temporal mapping &
\(\mathcal S_{\mathrm{sem},t}^{\mathcal T}=f(\mathcal P_{phy,t-W+1},\ldots,\mathcal P_{phy,t})\)\par
Summarize motion over time for gesture recognition~\cite{amin2019hand}. \\
\specialrule{0.3pt}{2pt}{2pt}
Contextual mapping &
\(\mathcal S_{\mathrm{sem}}^{\mathcal T}=\{f(\mathbf r_l,\mathcal R_j)\}_{l\in\mathcal I_{\mathcal T},\,j=1,\ldots,J}\)\par
Determine region membership for intrusion detection~\cite{mojahedian2026task}. \\
\specialrule{0.3pt}{2pt}{2pt}
Learned mapping &
\(\mathcal S_{\mathrm{sem}}^{\mathcal T}=\mathcal M_{\psi}(\mathcal P_{phy}\mid\mathcal T)\)\par
Learn a semantic representation for all listed applications. \\
\bottomrule
\end{tabular}
\end{table}

\section{Semantic Sensing Framework}
\label{sec:semantic_sensing_framework}
This section presents the SemS framework, detailing its core architecture and transceiver co-design. We provide an information-theoretic interpretation of the objective for task-relevant information extraction under practical resource constraints.
\subsection{Architecture and Main Components}
\label{sec:sems_architecture}

\begin{figure}[t]
    \centering
    \includegraphics[width= 1 \linewidth]{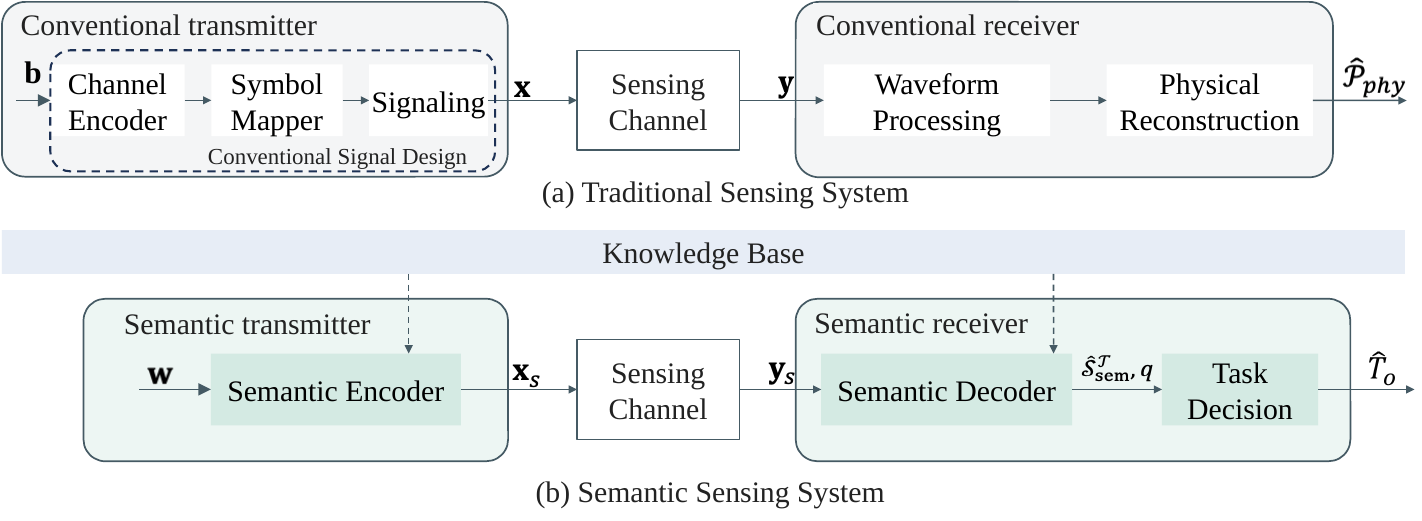}
    \caption{The proposed SemS system.}
    \label{fig:SS_frame}
\end{figure}

Fig. \ref{fig:SS_frame} compares conventional reconstruction-oriented sensing with SemS in a digital baseband model. The conventional transmitter uses a communication waveform as an example, mapping information bits $\mathbf b$ through channel encoding, symbol mapping, and signaling to the probing signal $\mathbf x$. Signaling includes digital waveform generation, such as OFDM modulation. The conventional receiver applies waveform processing, such as matched filtering or demodulation, followed by physical reconstruction. The SemS transceiver comprises the following functional modules.
{
\subsubsection{Knowledge Base (KB)} It provides task-aware priors and contextual constraints to the semantic transceiver. This background knowledge can be explicit, such as external knowledge graphs or databases, or implicit, learned from data and embedded within the parameters of trained models. Irrespective of its format, the KB can also offer prior information about interference to both the transmitter and receiver. In this implementation, the semantic encoder and decoder learn task- and environment-dependent knowledge from training data through offline joint optimization. Their parameters remain unchanged during inference, while the semantic decoder produces updated estimates from new observations. If the deployed model no longer meets task requirements under changed environmental statistics or tasks, representative new data can be used for offline fine-tuning or retraining.

\subsubsection{Semantic Encoder} Active probing in SemS uses transmit waveform design to enhance the observability of task-relevant information. The semantic encoder $f_{\boldsymbol{\Theta}_t}(\cdot)$ maps a digital input $\mathbf w$ to a task-aware probing signal $\mathbf x_s$. It can replace selected modules or the entire sequence within the dashed box in Fig. \ref{fig:SS_frame}. When the entire sequence is replaced, $\mathbf w$ represents the information-bit sequence. For partial replacement, the representation of $\mathbf w$ depends on the modules being replaced, with retained upstream processing supplying inputs such as coded bits, mapped symbols, or a baseline waveform. Like classical matched illumination \cite{martone2021view}, our strategy uses prior knowledge to guide waveform optimization for semantic recovery, with training data representing the target and environment statistics. For example, in a pedestrian-intent task, the encoder could be designed to improve the observability of motion-dependent features relevant to the intended decision.

\subsubsection{Sensing Channel} It models the physical interaction between the transmitted waveform and the environment through the LTV model in \eqref{eq:ltv_channel}.
Semantic recovery through this channel and the semantic decoder is characterized statistically in Sec.~\ref{sec:semantic_recovery}, while a more general explicit equivalent semantic channel is left for future work.

\subsubsection{Semantic Decoder} Using a parametric function $g_{\boldsymbol{\Theta}_r}(\cdot)$, this module estimates the semantic state $\hat{\mathcal S}_{\mathrm{sem}}^{\mathcal T}$ from the received baseband observation $\mathbf y_s$. It also provides the normalized predictive distribution $q_{\boldsymbol\Theta_r}(\cdot\mid\mathbf y_s)$, denoted by $q$. For example, a semantic decoder could estimate pedestrian intent from received echoes for downstream driving decisions.

\subsubsection{Task Decision} Decision rule $\mathcal D_T$ uses the decoder's predictive distribution and the task mapping $\mathcal M_T$ to obtain the task output $\hat T_o=\mathcal D_T[q_{\boldsymbol\Theta_r}(\cdot\mid\mathbf y_s)]$. This task-inference function can be implemented using a conventional algorithm or a neural network.\par
}

% %============The Main Components of SemS Systems =======

\subsection{Semantics-Oriented Transceiver Co-Design}
\label{sec:transceiver_codesign}

Within this SemS architecture, we jointly design the semantic encoder and decoder to acquire the information required by the sensing task. Let $\mathbf w$ denote the digital input supplied to the semantic encoder. The semantic encoder $f_{\boldsymbol{\Theta}_t}(\cdot)$, with trainable parameters $\boldsymbol\Theta_t$, maps this input to the probing signal in the selected digital baseband representation as
\begin{align}
    \mathbf{x}_s = f_{\boldsymbol{\Theta}_t} (\mathbf{w}) \in \mathbb{C}^{N}.
    \label{def_encoder}
\end{align}
Here, $N$ is the dimension of the output representation. The mapping $f_{\boldsymbol\Theta_t}$ includes the retained digital operations between the specified input and output interfaces, while $\boldsymbol\Theta_t$ parameterizes the modules optimized for semantic sensing. \par
% BEGIN R1-5; M017 receiver
Let $\mathbf H\in\mathbb C^{N_y\times N}$ denote the effective matrix representation of the discrete LTV channel in the chosen signal domain, including the specified linear transmit and receive processing between $\mathbf x_s$ and $\mathbf y_s$, where $N_y$ is the dimension of the received observation. For this processing configuration, $\mathbf H$ is determined by the physical path set, $\mathcal P_{phy}$. The received observation $\mathbf y_s\in\mathbb C^{N_y}$ is expressed as
{
\begin{equation}
\mathbf y_s=\mathbf H\mathbf x_s+\mathbf n,
\label{eq:physical_observation}
\end{equation}
}%
where $\mathbf n\in\mathbb C^{N_y}$ denotes additive receiver noise in the observation domain. All modeled target and clutter paths remain in $\mathbf H$.
Signal $\mathbf y_s$ is then processed by the semantic receiver, $g_{\boldsymbol\Theta_r}(\cdot)$, parameterized by $\boldsymbol\Theta_r$, which is designed to estimate task-relevant semantics directly from the received physical signal. The decoding process yields the estimated semantic state and the corresponding task output as
\begin{align}
\hat{\mathcal S}_{\mathrm{sem}}^{\mathcal T}=g_{\boldsymbol\Theta_r}(\mathbf y_s), \,\,
\hat T_o=\mathcal D_T[q_{\boldsymbol\Theta_r}(\cdot\mid\mathbf y_s)].
\label{def_decoder}
\end{align}
Here, $q_{\boldsymbol\Theta_r}(\cdot\mid\mathbf y_s)$ denotes the semantic decoder's normalized predictive distribution. Decision rule $\mathcal D_T$ uses this distribution and the task mapping $\mathcal M_T$ to obtain the task output.
% END R1-5; M017 receiver

The semantic representation $\mathcal S_{\mathrm{sem}}^{\mathcal T}$, defined in (\ref{eq:semantic_projection}), captures task-relevant information about the physical environment for downstream inference. It can take the form of a learned latent vector or an interpretable attribute vector, with the latter providing intermediate concepts for downstream prediction \cite{koh2020concept}. The task-level design objective for encoder parameters ${\boldsymbol{\Theta}_t}$ and decoder parameters ${\boldsymbol{\Theta}_r}$ is
\begin{align}
    (\boldsymbol{\Theta}_t^*, \boldsymbol{\Theta}_r^*) = \arg \min_{\boldsymbol{\Theta}_t, \boldsymbol{\Theta}_r} \mathbb{E} \Big[ \mathcal{L}_{\text{task}}(T_o, \hat{T}_o) \Big],
    \label{eq:task_level_objective}
\end{align}
where $\mathcal{L}_{\text{task}}(\cdot)$ is the loss function that quantifies the disparity between the ground truth and the prediction. In the OFDM implementation, we train the transceiver using the semantic-recovery loss and evaluate the task performance through the specified task mappings. More generally, sensing quality-of-service (QoS) metrics can be combined according to the downstream task to form a semantic metric that reflects the end-to-end utility of the sensing result \cite{wen2023task,lu2024semantic}.\par

\subsection{Semantic Recovery and Information Flow}
\label{sec:semantic_recovery}
The semantic objective guides waveform and receiver optimization to recover the semantic state directly from $\mathbf y_s$. For a given sensing configuration, combining \eqref{eq:physical_observation} and \eqref{def_decoder} gives
{
\begin{equation}
\hat{\mathcal S}_{\mathrm{sem}}^{\mathcal T}=g_{\boldsymbol\Theta_r}\!\left(\mathbf H\mathbf x_s+\mathbf n\right).
\label{eq:semantic_recovery}
\end{equation}
}%
For a given task, probing waveform, and decoder, semantic recovery is characterized by the conditional distribution
{
\small
\begin{equation}
p(\hat{\mathcal S}_{\mathrm{sem}}^{\mathcal T}\!\mid\!\mathcal S_{\mathrm{sem}}^{\mathcal T};\mathbf x_s)=\int p(\hat{\mathcal S}_{\mathrm{sem}}^{\mathcal T}\!\mid\!\mathcal P_{phy};\mathbf x_s)\, dP(\mathcal P_{phy}\!\mid\!\mathcal S_{\mathrm{sem}}^{\mathcal T}).
\label{eq:semantic_transition}
\end{equation}
}%
Here, $P(\mathcal P_{phy}\mid\mathcal S_{\mathrm{sem}}^{\mathcal T})$ specifies the distribution of physical realizations mapped to the same semantic state by $\mathcal M$. For each realization, $p(\hat{\mathcal S}_{\mathrm{sem}}^{\mathcal T}\mid\mathcal P_{phy};\mathbf x_s)$ describes the decoder outputs induced by receiver noise through \eqref{eq:semantic_recovery}. Averaging over these realizations yields the distribution of the recovered semantic state for a given true semantic state.

The following theorem characterizes this information flow through the chain rule and the data processing inequality (DPI).
\begin{theorem}
\label{thm:DPI}
For the task-conditioned semantic mapping $\mathcal S_{\mathrm{sem}}^{\mathcal T}=\mathcal M(\mathcal P_{phy}\mid\mathcal T)$ and a given probing design, the chain rule gives
\begin{align}
I(\mathcal P_{phy};\mathbf y_s)
=I(\mathcal S_{\mathrm{sem}}^{\mathcal T};\mathbf y_s)+I(\mathcal P_{phy};\mathbf y_s\mid\mathcal S_{\mathrm{sem}}^{\mathcal T}).
\end{align}
Moreover, receiver processing satisfies
\begin{align}
I(\mathcal S_{\mathrm{sem}}^{\mathcal T};\hat{\mathcal S}_{\mathrm{sem}}^{\mathcal T})
\leq I(\mathcal S_{\mathrm{sem}}^{\mathcal T};\mathbf y_s).
\end{align}
\end{theorem}
% BEGIN 2026-09-17 information-decomposition proof
{
\begin{IEEEproof}
For brevity, write $P=\mathcal P_{phy}$, $S=\mathcal S_{\mathrm{sem}}^{\mathcal T}$, and $Y=\mathbf y_s$. By the definition $S=\mathcal M(P\mid\mathcal T)$ for the given task, $I(S;Y\mid P)=0$, and the chain rule gives
\begin{align}
I(P;Y)=I(P,S;Y)=I(S;Y)+I(P;Y\mid S).
\end{align}
Receiver processing forms the Markov chain $S\to Y\to\hat{\mathcal S}_{\mathrm{sem}}^{\mathcal T}$, so the DPI gives the stated inequality.
\end{IEEEproof}
}%
% END 2026-09-17 information-decomposition proof
Building on Theorem \ref{thm:DPI}, we use the information gap $I(\mathcal P_{phy};\mathbf y_s)-I(\mathcal S_{\mathrm{sem}}^{\mathcal T};\mathbf y_s)$ to quantify the information about physical variations within the same semantic state under the task-conditioned semantic mapping $\mathcal M(\cdot\mid\cdot)$. Furthermore, the DPI \cite{cover1999elements} limits the semantic information retained after receiver processing to that available in the physical observation.

\textit{Remark.} Given the probing design and semantic mapping, if $I(\mathcal P_{phy};\mathbf y_s\mid\mathcal S_{\mathrm{sem}}^{\mathcal T})=0$, then $I(\mathcal P_{phy};\mathbf y_s)=I(\mathcal S_{\mathrm{sem}}^{\mathcal T};\mathbf y_s)$. In this case, the observation contains no additional information about physical variations within the same semantic state.
% END R1-4/R1-5; M017/M018 model dependencies

% BEGIN 2026-09-15 information-theoretic interpretation
To relate the information in the observation to task performance, let $\hat T_o$ denote an estimate of task output $T_o$ obtained from $\mathbf y_s$. Given the probing design and the semantic and task mappings, let $D_T=\mathbb E[d_T(T_o,\hat T_o)]$ denote the achieved task distortion and $R_T(D)$ the rate-distortion function of $T_o$ under the task-specific distortion measure $d_T$ \cite{cover1999elements}. Since $T_o=\mathcal M_T(\mathcal S_{\mathrm{sem}}^{\mathcal T})$, the rate-distortion definition and DPI give
\begin{align}
R_T(D_T)&\leq I(T_o;\hat T_o)\leq I(T_o;\mathbf y_s)\leq I(\mathcal S_{\mathrm{sem}}^{\mathcal T};\mathbf y_s).
\end{align}
To achieve task distortion $D_T$, the observation $\mathbf y_s$ must contain sufficient information about task output $T_o$, as required by $I(T_o;\mathbf y_s)\geq R_T(D_T)$.\par
% END 2026-09-15 information-theoretic interpretation

\subsection{Relevance-Oriented Design Criterion}
\label{sec:design_criterion}
% BEGIN M018; R1-4/R1-5 information dependencies
The information decomposition in Theorem~\ref{thm:DPI} motivates a sensing criterion that prioritizes task-relevant information under sensing-resource constraints.
\subsubsection{Limitations of Conventional Transceivers}
\label{sec_metric_limitation}
While SemS prioritizes the recovery of $\mathcal{S}_{\text{sem}}^{\mathcal{T}}$, reconstruction-oriented ISAC transceivers evaluate performance through fidelity-based metrics, typically the ambiguity function (AF) for radar sensing and mutual information (MI) for channel estimation \cite{andrew2021overview}. However, using these metrics as the sole design objective can lead to a misalignment between physical fidelity and semantic effectiveness. \par
In classical radar theory, the AF characterizes the correlation between the transmitted waveform and its delayed and Doppler-shifted replicas. Full-scene resolution designs aim to shape the AF to approach an ideal ``thumbtack'' response, minimizing sidelobes to distinguish all point scatterers in the delay--Doppler domain. From an information-theoretic perspective, a separate sensing criterion measures the reduction in uncertainty regarding the global channel state $\mathbf{h}$ as
\begin{align}
I(\mathbf{h}; \mathbf{y} | \mathbf{x}) = H(\mathbf{y} | \mathbf{x}) - H(\mathbf{y} | \mathbf{x}, \mathbf{h}),
\end{align}
where $H(\cdot)$ represents entropy, with differential entropy used for continuous variables.\par
Both full-scene AF shaping and channel-information maximization emphasize the acquisition of physical information. However, Theorem \ref{thm:DPI} shows that this information can also reflect physical variations within the same semantic state. An increase in physical information therefore does not necessarily imply an increase in semantic information. Similar to semantic communication designs that incorporate latency and transmit-power constraints \cite{yang2023energy}, the SemS framework can be extended to include conventional physical sensing metrics as constraints or regularization terms. However, optimizing full-scene physical fidelity alone may direct sensing resources toward estimating task-irrelevant parameters, such as random phase rotations and multipath fading from non-target objects. \par
Beyond metric mismatch, reconstruction-oriented ISAC systems can optimize transmit waveforms according to physical-fidelity criteria independently of the sensing tasks at the receiver. Such decoupled designs may allocate limited power and bandwidth to estimating physical parameters that are irrelevant to $\mathcal{S}_{\text{sem}}^{\mathcal{T}}$.

\subsubsection{Traditional Sensing Formulation}
As illustrated in Fig. \ref{fig:SS_frame}, reconstruction-oriented sensing systems aim to estimate the modeled physical state, $\mathcal P_{phy}$, from noisy observations. The transmitter emits a probing waveform, which interacts with the physical environment to produce the received signal, and the receiver estimates the physical parameters from this observation. From an information-theoretic perspective, the acquisition of physical information can be quantified by the mutual-information criterion in Section~\ref{sec:design_criterion}.

\subsubsection{SemS Formulation}
In contrast, SemS shifts the design objective from reconstruction to relevance. As established in Theorem \ref{thm:DPI}, the physical information in the received observation includes both information about the semantic state and information about physical variations within the same semantic state. This distinction motivates waveform design that prioritizes the acquisition of task-relevant information. Accordingly, the proposed sensing formulation adopts this relevance objective under explicit waveform and sensing-resource constraints. For the semantic mapping defined in (\ref{eq:semantic_projection}), the goal is to maximize the mutual information between the observation $\mathbf y_s$ and the semantic state $\mathcal S_{\mathrm{sem}}^{\mathcal T}$, formulated as
\begin{align}
\max_{\boldsymbol\Theta_t}\quad &I(\mathcal S_{\mathrm{sem}}^{\mathcal T};\mathbf y_s)\nonumber\\
\mathrm{s.t.}\quad &\mathbf x_s=f_{\boldsymbol\Theta_t}(\mathbf w)\in\Omega,
\label{infor_IB}
\end{align}
where $\Omega$ specifies the admissible waveform designs under the available sensing resources. Within this feasible set, the waveform is optimized to maximize the semantic information available in the received observation. For direct task inference, $\mathcal S_{\mathrm{sem}}^{\mathcal T}=T_o$. Section IV develops a variational training objective for the resulting transceiver design.
% END M018; R1-4/R1-5 information dependencies
\begin{figure*}[tbh]
    \centering
    \includegraphics[width=0.8  \linewidth]{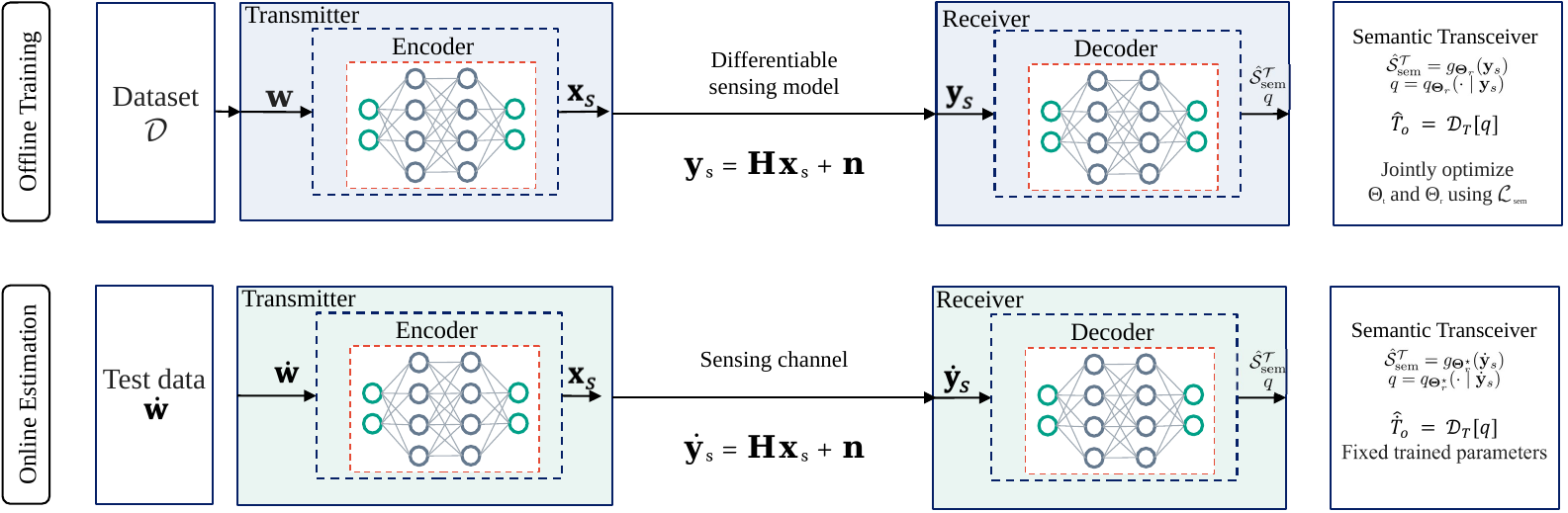}
    \caption{The proposed SemS-based end-to-end training framework. The upper portion illustrates offline transceiver training using labeled data, while the lower portion depicts online estimation using the optimized parameters.}
    \label{fig:SS_endtoend_framework}
\end{figure*}

\section{DL-based Optimization Framework}
% BEGIN M017/M018 metric dependencies
Building on the information-theoretic criterion in Section~\ref{sec:design_criterion}, this section develops a DL-based framework for jointly optimizing the transceiver to recover task-relevant semantic information under sensing-resource constraints. We first introduce metrics for evaluating physical reconstruction, semantic recovery, and task performance. We then formulate a differentiable signal propagation model and derive a variational lower bound for end-to-end training.
\subsection{Performance Metrics}
The proposed SemS framework extends direct task-oriented inference by explicitly describing the task-relevant environmental information used to produce the task output. This information is represented by $\mathcal S_{\mathrm{sem}}^{\mathcal T}$, and decision rule $\mathcal D_T$ obtains $\hat T_o$ from the decoder posterior. Accordingly, we evaluate physical reconstruction, recovery of this semantic representation, and the resulting task performance while retaining the task-oriented optimization objective.
\subsubsection{Physical Space}
In reconstruction-oriented sensing, physical metrics, such as channel MSE, characterize the fidelity of the recovered physical information. Within the SemS framework, these metrics quantify physical reconstruction accuracy, while waveform design is subject to the sensing-resource constraints. As discussed in Section~\ref{sec:design_criterion}, conventional physical metrics can also be retained as constraints or regularization terms when baseline physical performance is required.
\subsubsection{Semantic Space}
In the semantic space, we evaluate how accurately the task-relevant environmental information represented by $\mathcal S_{\mathrm{sem}}^{\mathcal T}$ is recovered. The choice of distortion measure depends on the form of this representation. For an interpretable attribute vector, one possible choice is a normalized form of the component-wise concept loss in \cite{koh2020concept}, $D_{\mathrm{sem}}\!\left(\mathcal S_{\mathrm{sem}}^{\mathcal T},\hat{\mathcal S}_{\mathrm{sem}}^{\mathcal T}\right)=\frac{1}{J}\sum_{j=1}^{J}d_j(s_j,\hat s_j)$, where $J$ is the number of attributes, $s_j$ and $\hat s_j$ are the true and estimated values of attribute $j$, respectively, and $d_j$ measures the corresponding recovery error. Depending on the attribute, $d_j$ can be a classification error for a categorical property or a squared error for a continuous quantity. For example, in human activity recognition, these attributes can describe body posture and limb motion, while $\mathcal M_T$ maps these attributes to an activity label such as walking.

For a learned latent representation in which vector direction reflects task-relevant similarity, cosine dissimilarity provides another possible measure \cite{qin2021semantic}. It is given by $D_{\mathrm{cos}}(\mathbf s,\hat{\mathbf s})=1-\frac{\mathbf s^{\mathsf T}\hat{\mathbf s}}{\|\mathbf s\|_2\|\hat{\mathbf s}\|_2}$, where $\mathbf s$ and $\hat{\mathbf s}$ denote the nonzero reference and recovered semantic vectors in the same embedding space. This measure evaluates their directional agreement and is invariant to positive rescaling of either vector. For example, when a motion representation encodes the activity pattern in its direction, cosine dissimilarity measures how closely the recovered vector preserves that direction.
\subsubsection{Task Space}
Decision rule $\mathcal D_T$ uses the decoder posterior to obtain the final task output $\hat T_o$. Task performance is then evaluated by comparing this output with the ground truth $T_o$ using a task-specific loss,
{
\begin{equation}
\begin{aligned}
\mathcal L_{\mathrm{task}}(T_o,\hat T_o)=\mathcal L_{\mathrm{task}}\!\left(\mathcal M_T(\mathcal S_{\mathrm{sem}}^{\mathcal T}),\mathcal D_T[q_{\boldsymbol\Theta_r}(\cdot\mid\mathbf y_s)]\right).
\end{aligned}
\label{eq:task_space_loss}
\end{equation}
}%
\noindent For classification tasks, such as activity recognition, a common training loss is cross-entropy (CE) \cite{shao2021learning}, given by $\mathcal L_{\mathrm{CE}}=-\sum_{k=1}^{K}y_k\log\hat y_k$, where $K$ denotes the number of task classes, $y_k$ is the ground-truth class indicator, and $\hat y_k$ is the predicted class probability. CE is evaluated on the predicted class probabilities before the final decision. In this example, the loss penalizes a low predicted probability for the correct activity label.

For regression tasks with continuous outputs, squared error is a common choice of task loss \cite{koh2020concept}. The corresponding task MSE is $\mathrm{MSE}_{\mathrm{task}}=\mathbb E\!\left[\|T_o-\hat T_o\|_2^2\right]$, where the expectation is taken over the joint distribution of the true and predicted task outputs. For example, in target localization, $T_o$ and $\hat T_o$ represent the true and estimated target positions, so this metric measures the average squared position error. When $\mathcal M_T$ is the identity mapping, the semantic state and task output coincide, and the same metrics apply to both. For example, in target ranging, the continuous target distance can be chosen as both the semantic state $\mathcal S_{\mathrm{sem}}^{\mathcal T}$ and the task variable $T_o$. The task mapping leaves this distance unchanged, so the recovered semantic state directly provides the range estimate.

% BEGIN R2-5 manuscript remark only
For a specified sensing task and configuration, analytical modeling or empirical calibration can estimate the expected semantic loss as a function of physical performance metrics. This relationship can then identify the metric values required to meet a specified semantic-loss target.
% END R2-5 manuscript remark only
% END M017/M018 metric dependencies

\subsection{DL-based End-to-End Optimization Framework}
This subsection presents a unified end-to-end optimization framework for SemS systems. We adopt the signal model in (\ref{def_encoder}), i.e., $\mathbf{x}_s = f_{\boldsymbol{\Theta}_t} (\mathbf{w}) \in \Omega$, where \(\Omega\) denotes the feasible waveform set under constraints such as amplitude and power limits. Joint training optimizes the selected transmitter modules through $\boldsymbol\Theta_t$ and the semantic decoder through $\boldsymbol\Theta_r$. The observation follows the baseband LTV model in Section~\ref{sec:physical_sensing_model}, as given in \eqref{eq:physical_observation}. Based on this discrete observation, the semantic decoder recovers $\hat{\mathcal S}_{\mathrm{sem}}^{\mathcal T}=g_{\boldsymbol\Theta_r}(\mathbf y_s)$, and the decision rule yields $\hat T_o=\mathcal D_T[q_{\boldsymbol\Theta_r}(\cdot\mid\mathbf y_s)]$, as defined in (\ref{def_decoder}). \par

% BEGIN M018; R1-5 training dependencies
As defined in (\ref{infor_IB}), the sensing objective is to maximize the mutual information between the semantic state and the received observation. The receiver subsequently recovers the semantic state through the decoder in (\ref{def_decoder}), giving the conditional distribution in (\ref{eq:semantic_transition}). To jointly design the waveform and receiver, we seek to maximize the mutual information between the true semantic state and its decoder estimate.
{
\begin{align}
\max_{\boldsymbol\Theta_t,\boldsymbol\Theta_r}\quad
&I(\mathcal S_{\mathrm{sem}}^{\mathcal T};\hat{\mathcal S}_{\mathrm{sem}}^{\mathcal T})\nonumber\\
\mathrm{s.t.}\quad
&\hat{\mathcal S}_{\mathrm{sem}}^{\mathcal T}=g_{\boldsymbol\Theta_r}(\mathbf y_s),\quad
\mathbf x_s=f_{\boldsymbol\Theta_t}(\mathbf w)\in\Omega,\nonumber\\
&{\mathbf y_s=\mathbf H\mathbf x_s+\mathbf n.}
\label{genral_optimization}
\end{align}
}%
\noindent By Theorem \ref{thm:DPI}, the semantic information retained in the decoder output is bounded by that available in the received observation. For tractable joint transceiver training, we use a variational lower bound on the mutual information between the semantic state and the received observation as a surrogate objective for semantic recovery. This bound uses the decoder's model of the conditional distribution of the semantic state given the observation, as described below \cite{barber2004algorithm}.
\begin{theorem}
For random variables $X,Y$ and a normalized variational conditional distribution $q(x\mid y)$, assume that the displayed quantities are finite. Then
{
\begin{align}
I(X;Y)\geq H(X)+\mathbb E_{p(x,y)}[\log q(x\mid y)].
\end{align}
}%
\noindent The gap is $\mathbb E_{p(y)}D_{\mathrm{KL}}(p(x\mid y)\|q(x\mid y))$, where $D_{\mathrm{KL}}$ denotes the Kullback--Leibler divergence, and $p(x,y)$, $p(y)$, and $p(x\mid y)$ are the true joint, marginal, and conditional distributions, respectively.
\label{theorem2}
\end{theorem}
Building upon Theorem \ref{theorem2}, we establish a tractable objective for the proposed framework. Substituting the semantic state and the received observation for $X$ and $Y$, respectively, gives
\begin{align}
I(\mathcal S_{\mathrm{sem}}^{\mathcal T};\mathbf y_s)
\geq H(\mathcal S_{\mathrm{sem}}^{\mathcal T}) +\mathbb E[\log q_{\boldsymbol\Theta_r}(\mathcal S_{\mathrm{sem}}^{\mathcal T}\mid\mathbf y_s)].
\label{throrem_2}
\end{align}
During transceiver training, the variational posterior is parameterized by the semantic receiver, while the semantic mapping and the distribution of physical parameters remain unchanged. Consequently, the semantic-state entropy is independent of the trainable transceiver parameters, so maximizing the lower bound is equivalent to maximizing its expected log-likelihood term. We therefore define the semantic training loss as
{
\begin{align}
\mathcal L_{\mathrm{sem}}=-\mathbb E[\log q_{\boldsymbol\Theta_r}(\mathcal S_{\mathrm{sem}}^{\mathcal T}\mid\mathbf y_s)].
\label{loss_variable}
\end{align}
}%
\noindent The decoder posterior $q_{\boldsymbol\Theta_r}$ defines the semantic training loss in (\ref{loss_variable}), which reduces to CE for classification \cite{shao2021learning}.
% END M018; R1-5 training dependencies

\subsection{Training and Deployment Strategy}
Building upon the differentiable optimization framework, we adopt a data-driven strategy comprising offline training and online deployment phases, as illustrated in Fig. \ref{fig:SS_endtoend_framework}.
\subsubsection{Training Phase}
% BEGIN M018; R1-4/R1-5 training dependencies
We utilize training samples containing physical channel realizations and their corresponding semantic labels. For each sample, the encoder generates the probing waveform, the physical channel model generates the noisy observation, and the receiver estimates the task-relevant semantic state. The waveform belongs to set $\Omega$ defined by the power and sensing-resource constraints. Incorporating this constraint into the semantic training objective gives
{
\begin{align}
\min_{\boldsymbol\Theta_t,\boldsymbol\Theta_r}\quad
&-\mathbb E\!\left[\log q_{\boldsymbol\Theta_r}\!\left(
\mathcal S_{\mathrm{sem}}^{\mathcal T}\mid
{\mathbf H f_{\boldsymbol\Theta_t}(\mathbf w)+\mathbf n}
\right)\right]\nonumber\\
\mathrm{s.t.}\quad & f_{\boldsymbol\Theta_t}(\mathbf w)\in\Omega.
\label{eq:objective}
\end{align}
}%
\noindent Here, the expectation is taken over the physical parameters and receiver noise. The resource constraint is imposed on the designed waveform. For direct estimation of a continuous task variable, squared error corresponds to a Gaussian variational posterior with constant variance \cite{barber2004algorithm}. Gradient-based methods have also been applied to constrained beamforming design in reconfigurable intelligent surface (RIS)-assisted systems \cite{yang2024blockage}. To enable end-to-end optimization, we implement the physical channel model as a differentiable layer within the computational graph.
% END M018; R1-4/R1-5 training dependencies

\subsubsection{Test Phase} During the test phase, given the optimized parameters, $(\boldsymbol{\Theta}_{\mathrm{t}}^{\star},\boldsymbol{\Theta}_{\mathrm{r}}^{\star})$, and a test input $\dot{\mathbf{w}}$, the input passes through the encoder and the physical channel. At the receiver, the received observation, $\dot{\mathbf{y}}_s$, is processed by the decoder network, whose posterior is used by decision rule $\mathcal D_T$ to obtain the predicted task output $\hat{\dot{T}}_o$. When the waveform or channel model changes, the transceiver is retrained or fine-tuned under the new configuration. A specific implementation is presented in Section \ref{secV}.\par

\section{Task-oriented Pilot Design}
\label{secV}

% BEGIN OFDM realization figure for author review
\begin{figure*}[t]
    \centering
    \includegraphics[width=0.8\textwidth]{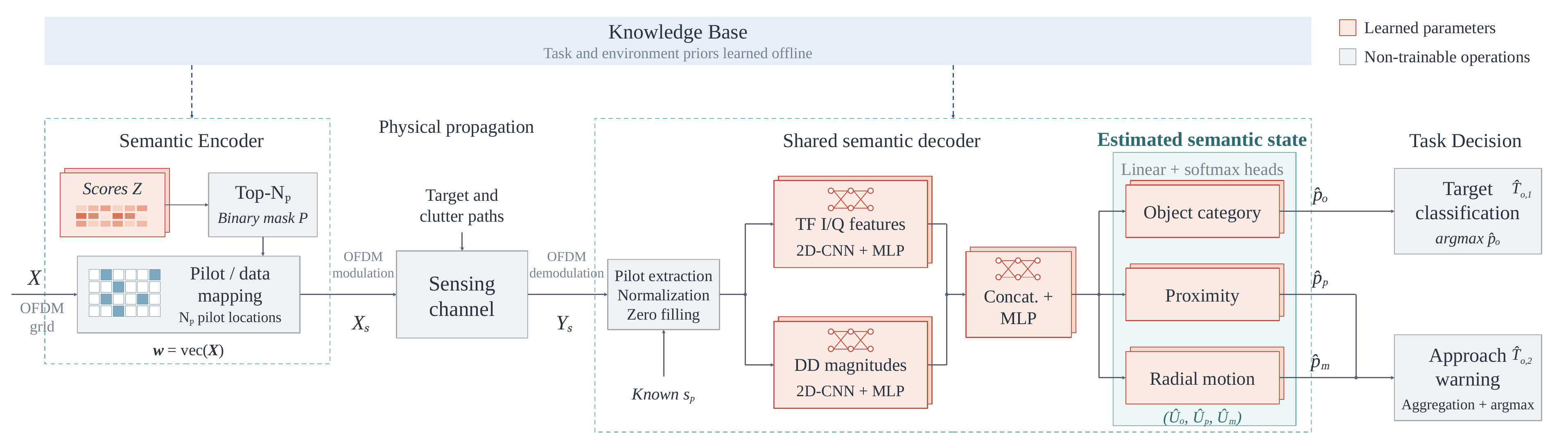}
    \caption{OFDM realization of SemS for target classification and approach warning.}
    \label{fig:ofdm_sems_realization}
\end{figure*}
% END OFDM realization figure for author review

To validate the efficacy of the proposed framework, we instantiate the SemS architecture within a single-antenna OFDM sensing system. In this example, pilot selection and receiver processing are jointly optimized to recover a shared semantic representation that supports target classification and approach warning through predefined task mappings. This section details the physical channel interaction within the OFDM-based SemS framework, formulates the transceiver optimization problem, and presents the corresponding DL-based solution.
\subsection{OFDM-based SemS System}

 We define a frame consisting of $N$ time slots and $M$ subcarriers, with $\mathbf X\in\mathbb C^{N\times M}$ denoting the data-symbol grid. In this realization, the encoder optimizes pilot placement using $\mathbf X$ as its input, with the data symbols and OFDM modulation specified. Thus, $\mathbf w=\operatorname{vec}(\mathbf X)$, where $\operatorname{vec}(\cdot)$ denotes matrix vectorization. The encoder constructs the time--frequency (TF) grid $\mathbf{X}_{\text{s}}=f_{\boldsymbol{\Theta}_{\text{t}}}(\mathbf{X}) \in \mathbb{C}^{N \times M}$, whose vectorized form is $\mathbf x_s=\operatorname{vec}(\mathbf X_s)$. This matrix mapping represents the encoder in (\ref{def_encoder}) in the TF domain. Standard OFDM modulation and cyclic prefix (CP) insertion at the transmitter, followed by CP removal and OFDM demodulation at the receiver, give the TF-domain input--output relationship
\begin{align}
\mathbf{Y}_{\mathrm{s}} = \mathbf{H}_{\mathrm{tf}} \odot \mathbf{X}_{\mathrm{s}} + \mathbf{V}_{\mathrm{tf}},
\label{ofdm_SemS}
\end{align}
where $\mathbf Y_s\in\mathbb C^{N\times M}$ is the received TF grid, $\mathbf{H}_{\mathrm{tf}}\in\mathbb{C}^{N\times M}$ represents the effective time–frequency channel response, and $\odot$ denotes the element-wise Hadamard product. Its vectorized form is $\mathbf y_s=\operatorname{vec}(\mathbf Y_s)$. % BEGIN R1-4 OFDM noise
{
Under ideal synchronization and negligible inter-carrier interference (ICI), $\mathbf V_{\mathrm{tf}}$ represents additive receiver noise, while all modeled target and clutter paths remain in $\mathbf H_{\mathrm{tf}}$.
}
% END R1-4 OFDM noise

The $(n,m)$-th element of $\mathbf{H}_{\mathrm{tf}}$ is modeled as \cite{andrew2021overview}
{\small
\begin{align}
    H_\mathrm{tf}[n,m]=\sum_{l=1}^L\alpha_lA_{g_\mathrm{rx},g_\mathrm{tx}}(\nu_l,\tau_l)e^{j2\pi\nu_lnT_\mathrm{sym}}e^{-j2\pi m\Delta f\tau_l},
\end{align}}%
where $L$ is the number of paths, and $\alpha_l$, $\tau_l$, and $\nu_l$ denote the complex reflection coefficient, delay, and Doppler shift of the $l$-th path, respectively. The slow-time sampling interval is $T_{\mathrm{sym}}$, and $\Delta f$ is the subcarrier spacing. The ambiguity function $A_{g_{\mathrm{rx}},g_{\mathrm{tx}}}(\cdot)$ is defined as
\begin{align}
A_{g_{\mathrm{rx}},g_{\mathrm{tx}}}(\nu,\tau) = \int g_{\mathrm{rx}}(t) g^{*}_{\mathrm{tx}}(t-\tau) e^{-j2\pi\nu t} dt,
\end{align}
where $g_{\mathrm{tx}}(t)$ and $g_{\mathrm{rx}}(t)$ denote the transmit pulse-shaping function and the receive filter, respectively. Assuming that the CP length exceeds the maximum delay spread and the pulse shaping is ideal, the ambiguity function can be approximated as unity. Furthermore, to capture the fine-grained features of targets, we incorporate the micro-Doppler effect into the phase term. The channel response is thus refined as \cite{wang2025bistatic}
\begin{align}
    H_{\mathrm{tf}}[n,m] \approx \nonumber  \sum_{l=1}^L &\alpha_l e^{-j2\pi m \Delta f \tau_l} \cdot e^{j2\pi \nu_l n T_{\mathrm{sym}}} \\ &\cdot e^{j \frac{4\pi}{\lambda} A_{l} \sin(\omega_{l} n T_{\mathrm{sym}} + \phi_l)},
    \label{eq:channel_micro_doppler}
\end{align}
where $\lambda$ is the carrier wavelength, and $A_l$, $\omega_l$, and $\phi_l$ denote the equivalent spatial displacement, dominant angular frequency, and initial phase for the $l$-th path, respectively, effectively forming a first-order sinusoidal approximation of target micro-motion. {\color{black}In this single-antenna realization, angular terms are omitted, and the reduced state $[\tau_l,\nu_l,\alpha_l]^T$ is augmented with the micro-motion parameters above. The resulting path set, \mbox{$\mathcal P_{phy}=\{(\alpha_l,\tau_l,\nu_l,\omega_l,A_l,\phi_l)\}_{l=1}^{L}$}, specifies the parameters of all modeled target and clutter paths.} % BEGIN R1-4/R1-5 OFDM physical observation
Under the stated OFDM assumptions, \eqref{ofdm_SemS} expresses the discrete LTV observation model in \eqref{eq:physical_observation} in the TF domain.\par
% END R1-4/R1-5 OFDM physical observation
% 等效信道，物理信道，但是可以表示成一个等效信道。变成了什么？在语义层面的关系。会不会是一个线性的关系。
% 传统是线性和卷积的，在语义感知里面是变成了一个什么关系的，是一个线性相关还是的
% Knowledge base 
% 雷达不同的任务。
% 刘凡老师做的object相关的，优化内容，优化方法有很多种，不需要穷举。现有的优化方法都可以用上。把关系讲清楚。submision， constrain。
\subsection{Problem Formulation}
{
In general, the semantic representation is obtained through a task-conditioned mapping from physical information. In this OFDM case study, we use a simplified, manually defined shared representation $\mathcal S_{\mathrm{sem}}^{\mathcal T}=(U_o,U_p,U_m)$ for target classification and approach warning, where $U_o$, $U_p$, and $U_m$ denote object category, proximity, and radial motion, respectively. Here, $\mathcal T$ specifies both tasks, and their respective mappings use the relevant attributes of the shared semantic state. The semantic labels are specified before training from the simulated target category, range, and radial velocity.
}

In this implementation, task-aware probing uses a static, input-independent pilot pattern optimized offline for semantic recovery and used unchanged during inference. The encoder combines this pilot pattern with the input data-symbol grid $\mathbf X$ to form the transmitted TF grid $\mathbf X_s$ as
\begin{align}
\mathbf{X}_s = f_{\boldsymbol{\Theta}_t}(\mathbf{X}) \triangleq \sqrt{E_p} s_p \mathbf{P}_{\boldsymbol{\Theta}_t} + \sqrt{E_d} (\mathbf{1} - \mathbf{P}_{\boldsymbol{\Theta}_t}) \odot \mathbf{X},
\end{align}
where $\mathbf{P}_{\boldsymbol{\Theta}_t} \in \{0,1\}^{N\times M}$ denotes the binary allocation matrix representing the optimized pilot pattern derived from the training set, and $\mathbf{1}$ is the all-ones matrix of identical dimensions. Scalar $s_p$ represents the deterministic pilot symbol, while $E_p$ and $E_d$ indicate the transmit energy allocated to the pilot and data components, respectively. This multiplexing strategy allocates a limited pilot budget $N_p$ for semantic recovery, with the active pilot locations defined by index set $\mathcal{P}_\mathrm{b}=\{(n,m):[\mathbf P_{\boldsymbol\Theta_t}]_{n,m}=1\}$. At the receiver, the semantic decoder extracts the pilot-only observations to construct subset $\mathbf{Y}_{\mathcal{P}_\mathrm{b}}=\{[\mathbf{Y}_s]_{n,m}:(n,m)\in \mathcal{P}_\mathrm{b}\}$. For target classification and approach warning, the goal is to jointly optimize the encoder and the decoder $g_{\boldsymbol{\Theta}_r}(\cdot)$ to recover their shared semantic state. Therefore, the optimization problem can be modeled as
{
\begin{align}
    \min_{\boldsymbol\Theta_t,\boldsymbol\Theta_r}\quad
    &-\mathbb E\!\left[\log q_{\boldsymbol\Theta_r}\!\left(\mathcal S_{\mathrm{sem}}^{\mathcal T}\mid\mathbf Y_{\mathcal P_\mathrm b}\right)\right],\nonumber\\
    \mathrm{s.t.}\quad &\mathbf X_s=f_{\boldsymbol\Theta_t}(\mathbf X),\quad
    \sum_{n,m}[\mathbf P_{\boldsymbol\Theta_t}]_{n,m}=N_p,\nonumber\\
    &\mathbf Y_s=\mathbf H_{\mathrm{tf}}(\mathcal P_{phy})\odot\mathbf X_s+\mathbf V_{\mathrm{tf}}.
    \label{loss_OFDM_E}
\end{align}
}%
\noindent The loss uses the probabilities assigned to the true semantic labels, following (\ref{loss_variable}). The task mappings convert the predicted attribute probabilities into task probabilities, from which the final decisions are obtained.

\subsubsection{Target Classification}
{
Target classification distinguishes object categories through their radar signatures \cite{amiri2020micro}. The task mapping selects the object-category attribute, giving $T_{o,1}=\mathcal M_{T_1}(\mathcal S_{\mathrm{sem}}^{\mathcal T})=U_o$.
}

\subsubsection{Approach Warning}
{
Motivated by collision warning based on target range and relative motion \cite{lee2005evaluation}, we define $T_{o,2}=\mathcal M_{T_2}(\mathcal S_{\mathrm{sem}}^{\mathcal T})$ by
\begin{align}
T_{o,2}=\begin{cases}
w(U_p), & U_m=\mathrm{Approaching},\\
\text{no warning}, & \text{otherwise},
\end{cases}
\label{eq:approach_warning_mapping}
\end{align}
where $w(\cdot)$ maps the proximity of an approaching target to a warning category. It assigns high warning to Near, low warning to Middle, and no warning to Far.
}%

\subsubsection{Target Range Estimation}
Target range estimation recovers the target body range $R$ at the start of the frame. For the single-task specification $\mathcal T_{\mathrm r}$ of range estimation, we choose the continuous semantic state $\mathcal S_{\mathrm{sem}}^{\mathcal T_{\mathrm r}}=R$. The task mapping is the identity, giving $T_{o,3}=\mathcal M_{T_3}(\mathcal S_{\mathrm{sem}}^{\mathcal T_{\mathrm r}})=R$, so that recovery of the semantic state directly yields the range estimate.

\begin{table}[tbp]
\caption{Network Configuration and Output Dimensions}
\label{tab:network_arch}
\centering
\renewcommand{\arraystretch}{1.1}
% 确保表格宽度自适应单栏
\resizebox{\columnwidth}{!}{
\begin{tabular}{c|l|c}
\hline
\textbf{Module} & \textbf{Layer / Operation} & \textbf{Output Dims.} \\
\hline
\hline
\multicolumn{3}{c}{\textbf{Semantic Encoder}} \\
\hline
\multirow{2}{*}[-0.5ex]{\textbf{Pilot Selector}}
 & Trainable Scores $\mathbf Z$ & $N\times M$ \\
 & \begin{tabular}[c]{@{}l@{}}Top-$N_p$ Selection\\and Binary Mask\end{tabular} & $N\times M$ \\
\hline
\hline
\multicolumn{3}{c}{\textbf{Shared Semantic Decoder}} \\
\hline
\textbf{TF Feature Branch}
 & \begin{tabular}[c]{@{}l@{}}TF I/Q $\rightarrow$\\2D-CNN + MLP\end{tabular} & $64$ \\
\textbf{DD Feature Branch}
 & \begin{tabular}[c]{@{}l@{}}DD Magnitudes $\rightarrow$\\2D-CNN + MLP\end{tabular} & $64$ \\
\textbf{Feature Fusion}
 & Concatenation + MLP & $128$ \\
\hline
\begin{tabular}[c]{@{}c@{}}\textbf{Semantic-Attribute}\\\textbf{Heads}\end{tabular}
 & Linear + Softmax & $3\times 3$ \\
\hline
\hline
\multicolumn{3}{c}{\textbf{Task Inference}} \\
\hline
\begin{tabular}[c]{@{}c@{}}\textbf{Target}\\\textbf{Classification}\end{tabular}
 & \begin{tabular}[c]{@{}l@{}}Object-Category\\Probabilities + Argmax\end{tabular} & $3\rightarrow1$ \\
\textbf{Approach Warning}
 & \begin{tabular}[c]{@{}l@{}}Probability Aggregation\\+ Argmax\end{tabular} & $3\rightarrow1$ \\
\hline
\end{tabular}
}
\end{table}

\subsection{DL-based Solution}
For target classification and approach warning, the pilot selector and shared semantic decoder in Fig.~\ref{fig:ofdm_sems_realization} implement the semantic encoder in (\ref{def_encoder}) and decoder in (\ref{def_decoder}), respectively. Pilot placement is optimized for semantic recovery under the specified pilot budget in (\ref{loss_OFDM_E}). The network configurations and output dimensions are summarized in Table~\ref{tab:network_arch}. Here, 2D-CNN denotes a two-dimensional convolutional neural network, and MLP denotes a multilayer perceptron. The inputs comprise TF in-phase and quadrature (I/Q) components and delay--Doppler (DD) magnitude maps. The network architecture, training procedure, and task-probability aggregation for this realization are detailed below. \par

\subsubsection{Semantic Encoder}

% BEGIN R1-9 pilot-selector implementation
{
The semantic encoder uses differentiable pilot selection \cite{soltani2020pilot}. A trainable score matrix $\mathbf Z\in\mathbb R^{N\times M}$ assigns one score to each TF location. The forward and inference passes select the $N_p$ largest scores to form the binary mask $\mathbf P_{\mathrm{h}}=\mathbf P_{\boldsymbol\Theta_t}$, giving exactly $N_p$ distinct pilot locations. During training, backpropagation uses the derivative of the temperature-controlled soft mask
\begin{align}
[\mathbf P_{\mathrm{s}}]_{n,m}
= N_p\frac{\exp(z_{n,m}/\bar\tau)}{\sum_{n',m'}\exp(z_{n',m'}/\bar\tau)}.
\end{align}
Here, $z_{n,m}$ is the $(n,m)$-th selection score, and $\bar\tau>0$ is the soft-mask temperature. During early training, a decaying bias toward pilot locations spread across the TF grid and random Gumbel perturbations are added to these scores. Inference uses only the learned scores. The straight-through mask \cite{jang2017categorical} is $\mathbf P_{\mathrm{h}}-\operatorname{sg}(\mathbf P_{\mathrm{s}})+\mathbf P_{\mathrm{s}}$, where $\operatorname{sg}(\cdot)$ stops gradient propagation.
}
% END R1-9 pilot-selector implementation

\subsubsection{Semantic Decoding and Task Inference}
At the receiver, the decoder estimates the attribute probabilities of the shared semantic state $\mathcal S_{\mathrm{sem}}^{\mathcal T}=(U_o,U_p,U_m)$ from $\mathbf{Y}_{{\mathcal{P}}_\mathrm{b}}$ for downstream task inference. Each received pilot sample is normalized by the known transmitted pilot, yielding $\tilde h_{n,m}=[\mathbf Y_s]_{n,m}/(\sqrt{E_p}s_p)$ for $(n,m)\in\mathcal P_{\mathrm b}$. The input tensor is constructed by stacking the real and imaginary parts of the channel matrix, $\mathbf{\tilde{H}}$, as $\mathbf{T}_{\mathrm{in}}=\{\Re(\tilde{\mathbf{H}}),\Im(\tilde{\mathbf{H}})\}$, where $\tilde{\mathbf{H}}$ is a sparse matrix with entries $\tilde{h}_{n,m}$ at pilot locations and zeros elsewhere. \par
The shared semantic decoder uses two 2D-CNN branches to process the TF I/Q components and DD magnitude maps. The branch features are concatenated and fused by an MLP, then passed to three parallel semantic-attribute heads. Each head uses a softmax layer to estimate the probabilities of one semantic attribute. We denote the predicted attribute probabilities by $\hat p_a(u)$ for attribute $U_a$, where $a\in\{o,p,m\}$ and $u$ is a possible attribute value. The most probable value gives the attribute estimate $\hat U_a$ for evaluating semantic recovery.\par
{
The decoder approximates the joint posterior by the product of the three attribute distributions. Accordingly, (\ref{loss_OFDM_E}) is implemented as the sum of their CE losses. The selector and decoder are trained jointly, followed by decoder training without updating the selected pilot mask.\par
For each task, we aggregate the probabilities of semantic states assigned to each task category and then select the most probable category,
\begin{align}
\hat p_{T_{o,i}}(t)&=\sum_{\mathbf u:\,\mathcal M_{T_i}(\mathbf u)=t}\prod_{a\in\{o,p,m\}}\hat p_a(u_a),\nonumber\\
\hat T_{o,i}&=\arg\max_t\hat p_{T_{o,i}}(t),\qquad i\in\{1,2\}.
\label{eq:task_probability_readout}
\end{align}
Here, $\mathbf u=(u_o,u_p,u_m)$ denotes a semantic state, and $i=1$ and $i=2$ correspond to target classification and approach warning, respectively. For target classification, this aggregation reduces to $\hat p_{T_{o,1}}(t)=\hat p_o(t)$.
}\par

For range estimation, the receiver forms range profiles from the selected observations and their zero-sum combinations across slow-time snapshots. A residual one-dimensional CNN refines these profiles, and a softmax readout produces the range estimate as a probability-weighted average of candidate ranges. The pilot selector and receiver are jointly trained using absolute range error with auxiliary supervision of the range-response peak.\par

\begin{table}[bt]
\centering
\caption{Target Parameters in the Nominal Scenario}
\setlength{\tabcolsep}{5pt}
\begin{tabular}{@{}cccc@{}}
\toprule
\textbf{\shortstack{\strut Target\\[-3pt]\strut Class}} & \textbf{\shortstack{\strut Range\\[-3pt]\strut (m)}} & \textbf{\shortstack{\strut Moving Speed\\[-3pt]\strut (m/s)}} & \textbf{\shortstack{\strut Micro-motion\\[-3pt]\strut Frequency (Hz)}} \\
\midrule
Pedestrian & $[20,170]$ & $[1,2]$ & $[1,2]$ \\
Car & $[20,170]$ & $[1,20]$ & $[0,9.1]$ \\
Drone & $[20,170]$ & $[1,10]$ & $[80,120]$ \\
\bottomrule
\end{tabular}
\label{tab:range_velocity_micro}
\end{table}

\section{Simulation Results}
\label{sec_result}
In this section, we validate the effectiveness of the proposed SemS framework within an OFDM sensing system. Unless otherwise specified, each frame contains \(N=32\) slow-time slots and \(M=64\) subcarriers. The bandwidth is \(B_{\mathrm{W}}=30\,\mathrm{MHz}\), corresponding to a subcarrier spacing \(\Delta f= 468.75\,\mathrm{kHz}\). The carrier frequency is \(f_c=3\,\mathrm{GHz}\), and the wavelength is \(\lambda=c/f_c=0.1\,\mathrm{m}\) with \(c=3\times 10^{8}\,\mathrm{m/s}\). The slow-time sampling interval is $T_{\mathrm{sym}}=1\,\mathrm{ms}$. Each target consists of one body component and three coherent micro-motion components with relative range offsets in $[-0.04,0.04]$ m. The effective micro-motion amplitudes are sampled uniformly from $[0,0.4]$, $[0,0.35]$, and $[0,0.13]$ m, following the order of the three target classes in Table~\ref{tab:range_velocity_micro}, with independent phases in $[0,2\pi]$. The two static clutter paths have ranges within $[0,200]$ m. The complex reflection coefficients are drawn from circularly symmetric complex Gaussian distributions and normalized within the target and clutter components. The target response has unit average power, and two static clutter paths are scaled to give a signal-to-clutter ratio (SCR) uniformly distributed over $[-3,3]$ dB. The target parameters are generated using established radar sensing models \cite{molchanov2014classification}, with the class-specific kinematic settings for the nominal scenario detailed in Table~\ref{tab:range_velocity_micro}. The listed speeds are magnitudes for moving targets. Approaching and receding velocities are negative and positive, respectively, while quasi-static velocities lie in $[-0.3,0.3]$ m/s. Both pilot and data symbols are normalized to unit power, and the number of pilots is $N_p=32$. The normalized TF channel samples are corrupted by additive receiver noise $V_{\mathrm{tf}}[n,m]\sim\mathcal{CN}(0,\sigma_v^2)$, where $\sigma_v^2=P_{\mathrm{ref}}/10^{\mathrm{SNR}/10}$. For each channel realization, $P_{\mathrm{ref}}$ is the average channel power over the full TF grid.\par

For target classification and approach warning, we generate $32,400$ Monte Carlo frames equally distributed over the 27 semantic states and partition them into training, validation, and test sets in an \(8\!:\!1\!:\!1\) ratio. The test metrics are averaged over $3,240$ independent realizations for each operating point. Training SNRs are sampled uniformly over $[0,30]$ dB, and test SNRs range from 0 to 35 dB in 5 dB steps. The models are trained for 200 epochs with Adam using an initial learning rate of $0.001$, a weight decay of $10^{-4}$, and a batch size of 256. The selector temperature decreases linearly from 5.0 to 0.5 over the first 50 epochs, with a decrement of 0.09 per epoch. \par
The following benchmarks are considered for target classification and approach warning.
\begin{itemize}
    \item \text{Benchmark 1 (Full Grid).} The full TF grid is observed to provide a reference with complete sampling coverage. The noisy observations are fed to the semantic decoder to recover the three semantic attributes.
    \item \text{Benchmark 2 (Uniform).} Pilots are uniformly distributed across the TF grid \cite{soltani2020pilot}. The received pilot observations are used to recover the semantic attributes through the semantic decoder. As an extension, LMMSE + Classifier reconstructs the full TF channel from uniformly distributed pilots using a linear minimum mean-squared error (LMMSE) estimator \cite{tse2005fundamentals} with channel statistics estimated exclusively from the training set. The reconstructed channel is then classified using the Task-based receiver architecture.
    \item \text{Benchmark 3 (Random).} Pilot locations are randomly selected from the TF grid and remain unchanged during training and inference. The semantic decoder recovers the semantic attributes from the observations at these locations.
    \item \text{Benchmark 4 (MSE-based).} The pilot pattern and a channel reconstruction network are jointly optimized to minimize channel MSE \cite{soltani2020pilot}. The resulting pilot pattern is then used unchanged to train the semantic decoder for attribute recovery.
    \item \text{Benchmark 5 (Task-based).} For each task, a separate pilot pattern and receiver are jointly optimized using CE on the corresponding task labels \cite{huijben2020deep}. For classification, the receiver uses the same feature-extraction architecture as SemS and a single output head that predicts the target class. For approach warning, the receiver directly predicts the three warning classes.
\end{itemize}

The T1 pilots + shared decoder baseline uses the pilot pattern learned by the Task-based classifier and a separately trained semantic decoder for the three attributes, with the pilot locations kept unchanged. For approach warning, the Full Grid, Uniform, Random, and MSE-based benchmarks reuse their respective pilot patterns and semantic decoders, applying the task mapping in Section~\ref{secV} to obtain warning predictions.

\subsection{SemS-based Target Classification}
\label{sec_result_A}
In this subsection, we evaluate the performance of the proposed SemS framework on target classification. This task uses class-dependent motion and micro-motion information to distinguish the three target classes, while each target category spans the same combinations of proximity and radial-motion states. The object-category component of the shared semantic representation provides the task output.

\begin{figure}[t]
    \centering
    \begin{subfigure}{0.45\linewidth}
      \centering
      \includegraphics[width=\linewidth]{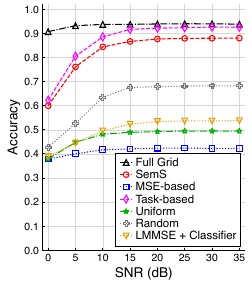}
      \caption{Accuracy in the nominal scenario.}
      \label{subfig_11_accuracy}
    \end{subfigure}
    \hfill
    \begin{subfigure}{0.45\linewidth}
      \centering
      \includegraphics[width=\linewidth]{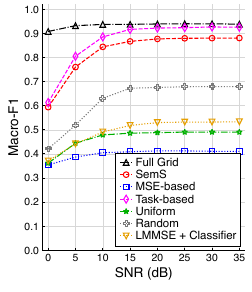}
      \caption{Macro-F1 in the nominal scenario.}
      \label{subfig_12_F1-Score}
    \end{subfigure}
    \vspace{1em}
    \begin{subfigure}{0.45\linewidth}
      \centering
      \includegraphics[width=\linewidth]{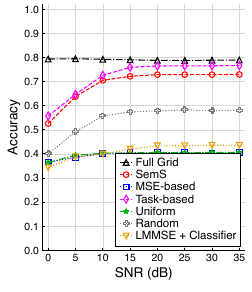}
      \caption{Accuracy in the overlap scenario.}
      \label{fig:overlap_accuracy}
    \end{subfigure}
    \hfill
    \begin{subfigure}{0.45\linewidth}
      \centering
      \includegraphics[width=\linewidth]{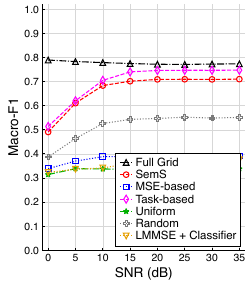}
      \caption{Macro-F1 in the overlap scenario.}
      \label{fig:overlap_macro_f1}
    \end{subfigure}
    \caption{Performance comparison with baselines for target classification in the nominal and overlap scenarios.}
    \label{Baselines1}
\end{figure}

To quantify the classification reliability, we employ two complementary metrics. Accuracy measures the overall correctness, while Macro-F1 gives equal weight to the F1 scores of all target classes \cite{shao2021learning}. Figs.~\ref{Baselines1}(\subref{subfig_11_accuracy}) and \ref{Baselines1}(\subref{subfig_12_F1-Score}) compare the performance over the tested SNR range. SemS consistently outperforms MSE-based, Uniform, and Random, while the Task-based classifier achieves higher performance and the Full Grid reference benefits from complete observations. Specifically, Uniform, Random, and MSE-based exhibit lower accuracy that saturates as SNR increases. Their performance gaps persist at high SNR, indicating that reducing observation noise alone does not recover the discriminative information obtained by semantic pilot design. The comparison with the MSE-based method illustrates the effect of the training objective under limited observations. Reconstruction loss evaluates the complete channel response, whereas the semantic loss emphasizes recovery of the three attributes used by the tasks. Both methods retain clutter in the physical channel, while their pilot-design objectives differ. The corresponding Macro-F1 trends in Fig.~\ref{Baselines1}(\subref{subfig_12_F1-Score}) also show a gain in class-averaged classification performance. The two-stage LMMSE baseline outperforms MSE-based, while its classification performance saturates below SemS at high SNR. The uniform pilot pattern used by LMMSE + Classifier provides temporally separated observations of target micro-motion, which can help distinguish the target categories. The higher performance of the Task-based classifier is consistent with its optimization for a single output, whereas SemS also recovers proximity and radial motion for the second task.

We consider a more challenging overlap scenario with moving-speed magnitudes of $[1,2]$, $[1,5]$, and $[1,4]$ m/s for the three target classes in Table~\ref{tab:range_velocity_micro}, retaining the same micro-motion models. Models trained on the nominal scenario are evaluated on independently generated overlap samples without retraining. In Figs.~\ref{Baselines1}(\subref{fig:overlap_accuracy}) and \ref{Baselines1}(\subref{fig:overlap_macro_f1}), SemS achieves higher accuracy and Macro-F1 than MSE-based, Uniform, Random, and LMMSE + Classifier throughout the tested SNR range. SemS and Random benefit from increasing SNR before reaching distinct plateaus, while Uniform and MSE-based show smaller gains and saturate at lower levels. The persistent gap between SemS and MSE-based supports selecting pilots for attribute recovery when radial velocity provides less separation between target classes. LMMSE + Classifier offers a modest accuracy improvement over Uniform and MSE-based at high SNR. Its Macro-F1 nevertheless remains below that of MSE-based, reflecting less balanced recovery across the classes. Task-based retains an advantage over SemS, with its pilot pattern and receiver optimized for classification alone, whereas SemS recovers the shared attributes used by both downstream tasks. Full Grid maintains the highest performance with complete TF observations, but its curves change little with SNR. The high-SNR plateaus show that reducing observation noise alone does not eliminate the remaining classification ambiguity.

\begin{figure}[tbp]
    \centering
    \begin{subfigure}[b]{0.45\linewidth}
        \centering
        \includegraphics[width=\linewidth]{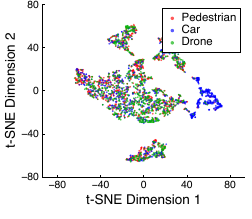}
        \caption{Uniform pilots.}
        \label{fig:sub1}
    \end{subfigure}
    \hfill 
    \begin{subfigure}[b]{0.45\linewidth}
        \centering
        \includegraphics[width=\linewidth]{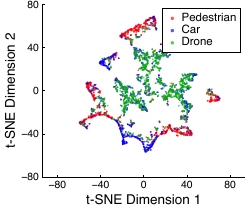}
        \caption{Random pilots.}
        \label{fig:sub2}
    \end{subfigure}
    
    \vspace{1em} 
    
    \begin{subfigure}[b]{0.45\linewidth}
        \centering
        \includegraphics[width=\linewidth]{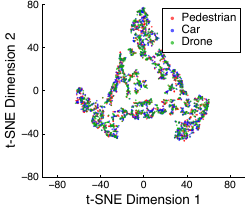}
        \caption{MSE-based pilots.}
        \label{fig:sub3}
    \end{subfigure}
    \hfill
    \begin{subfigure}[b]{0.45\linewidth}
        \centering
        \includegraphics[width=\linewidth]{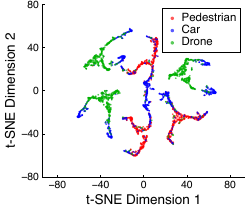}
        \caption{SemS pilots.}
        \label{fig:sub4}
    \end{subfigure}
    
    \caption{t-SNE visualization of semantic-attribute probabilities under different pilot patterns at 20 dB.}
    \label{fig:tsne_distributions}
\end{figure}

To provide deeper insight into transceiver co-design, we visualize the concatenated probabilities of the three semantic attributes using t-distributed stochastic neighbor embedding (t-SNE) \cite{vandermaaten2008visualizing} in Fig.~\ref{fig:tsne_distributions}. The four panels use the same test samples and color each sample by its target category. This projection qualitatively shows how samples from different target categories are separated in the decoder's predicted attribute probabilities. As observed in Fig.~\ref{fig:tsne_distributions}(\subref{fig:sub1}), Uniform produces substantial overlap between Pedestrian and Drone, although Car occupies more distinct regions. Random in Fig.~\ref{fig:tsne_distributions}(\subref{fig:sub2}) also exhibits mixed-category regions. Similarly, the MSE-based method in Fig.~\ref{fig:tsne_distributions}(\subref{fig:sub3}) produces several local groups with overlapping target categories. This pattern is consistent with its lower classification performance in Fig.~\ref{Baselines1}, where minimizing global reconstruction error does not provide equally effective target-category recovery. In contrast, SemS in Fig.~\ref{fig:tsne_distributions}(\subref{fig:sub4}) produces clearer regions dominated by individual target categories. A category can occupy several groups because the displayed representation also includes the other attribute probabilities. The t-SNE visualization provides qualitative support for the classification trends.

\begin{figure}[t]
    \centering
    % --- Subfigure 1: Lower clutter ---
    \begin{subfigure}{0.45\linewidth}
      \centering
      \includegraphics[width=\linewidth]{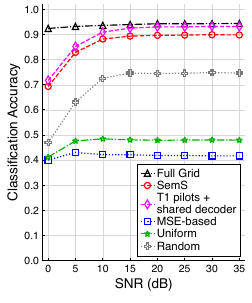}
      \caption{Lower clutter, SCR $=+6$ dB.}
      \label{fig:only_doppler}
    \end{subfigure}
    \hfill
    % --- Subfigure 2: Higher clutter ---
    \begin{subfigure}{0.45\linewidth}
      \centering
      \includegraphics[width=\linewidth]{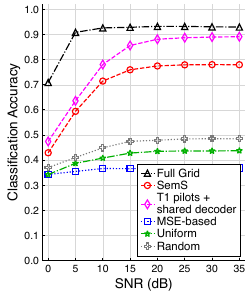}
      \caption{Higher clutter, SCR $=-6$ dB.}
      \label{fig: complex_channel}
    \end{subfigure}
    
    % --- Main Caption ---
    \caption{Target classification accuracy under different clutter levels.}
    \label{fig:extreme_cases}
\end{figure}

To evaluate performance under changes in clutter strength, we compare classification at lower and higher clutter levels outside the training range. Both conditions retain the target micro-Doppler components, and the pilot patterns and decoder parameters remain unchanged during evaluation. Fig.~\ref{fig:extreme_cases}(\subref{fig:only_doppler}) presents classification accuracy under lower clutter. SemS achieves higher accuracy than MSE-based, Uniform, and Random, while the shared decoder using T1-selected pilots and the Full Grid reference achieve higher accuracy than SemS. Random performs better than Uniform at moderate and high SNRs, showing that a regular sampling pattern does not necessarily preserve the most useful target information. In contrast, Fig.~\ref{fig:extreme_cases}(\subref{fig: complex_channel}) shows the effect of stronger clutter. The accuracy of all pilot-limited methods decreases, while Random remains better than Uniform. This overall degradation occurs because stronger clutter makes the target-dependent variations less prominent in the sampled channel response, increasing the difficulty of classification under the same pilot budget. SemS retains its advantage over Uniform, Random, and MSE-based, although its gap to Full Grid increases. These results support the robustness of semantic pilot design to the tested clutter changes while showing that sensing with limited observations remains sensitive to interference.\par

\begin{figure}[t]
    \centering
    % --- Subfigure 1: Warning accuracy ---
    \begin{subfigure}{0.45\linewidth}
      \centering
      \includegraphics[width=\linewidth]{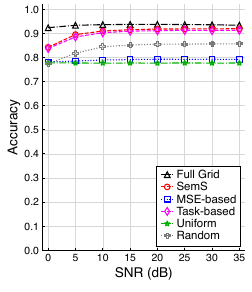}
      \caption{Accuracy.}
      \label{fig:warning_accuracy}
    \end{subfigure}
    \hfill
    % --- Subfigure 2: Warning Macro-F1 ---
    \begin{subfigure}{0.45\linewidth}
      \centering
      \includegraphics[width=\linewidth]{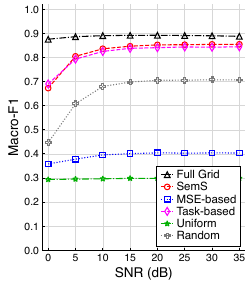}
      \caption{Macro-F1.}
      \label{fig:warning_macro_f1}
    \end{subfigure}
    
    \caption{Performance comparison with baselines for approach warning.}
    \label{Baselines2}
\end{figure}

\subsection{SemS-based Approach Warning}
\label{sec_result_B}
We next evaluate approach warning using the same SemS pilot pattern, decoder, and semantic representation as in Section~\ref{sec_result_A}. The predicted attribute probabilities are used to obtain warning levels through the task mapping defined in Section~\ref{secV}. Performance is evaluated using accuracy and Macro-F1, with Macro-F1 assigning equal weight to the three warning classes. \par
Fig.~\ref{Baselines2} illustrates the warning performance across varying SNRs. As shown by the accuracy curves in Fig.~\ref{Baselines2}(\subref{fig:warning_accuracy}), Uniform and MSE-based achieve relatively high overall correctness even at low SNR. Their substantially lower Macro-F1 in Fig.~\ref{Baselines2}(\subref{fig:warning_macro_f1}) reveals weaker recovery of the less frequent warning classes. SemS achieves higher Macro-F1 than MSE-based, Uniform, and Random throughout the tested SNR range. It also outperforms the separately trained Task-based warning model except at the lowest tested SNR, where SemS yields a slightly lower Macro-F1. These warning predictions use the same attribute outputs that support target classification, with only the predefined task readout changed.

\begin{figure}[t]
    \centering
    \begin{subfigure}{0.45\linewidth}
      \centering
      \includegraphics[width=\linewidth]{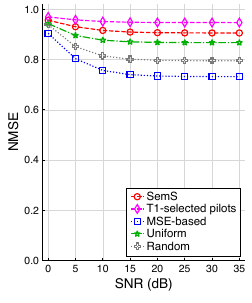}
      \caption{Channel reconstruction NMSE.}
      \label{fig:reconstruction_nmse}
    \end{subfigure}
    \hfill
    \begin{subfigure}{0.45\linewidth}
      \centering
      \includegraphics[width=\linewidth]{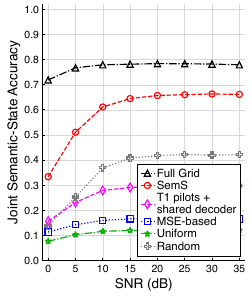}
      \caption{Joint semantic-state accuracy.}
      \label{fig:semantic_state_accuracy}
    \end{subfigure}
    \caption{Physical reconstruction and semantic recovery performance.}
    \label{fig:physical_semantic_performance}
\end{figure}

To examine the relation between physical fidelity and semantic recovery, Fig.~\ref{fig:physical_semantic_performance} compares the normalized mean-squared error (NMSE) of channel reconstruction with joint semantic-state accuracy. For the NMSE comparison in Fig.~\ref{fig:physical_semantic_performance}(\subref{fig:reconstruction_nmse}), we train a separate channel reconstruction network for each pilot pattern using the same architecture, training data, and training settings, while keeping the pilot locations unchanged. MSE-based achieves lower reconstruction NMSE than SemS in Fig.~\ref{fig:physical_semantic_performance}(\subref{fig:reconstruction_nmse}), whereas SemS achieves substantially higher joint semantic-state accuracy in Fig.~\ref{fig:physical_semantic_performance}(\subref{fig:semantic_state_accuracy}), which counts a sample as correct only when all three attributes are recovered correctly. The opposite ordering shows that a pilot pattern yielding lower reconstruction NMSE does not necessarily yield higher semantic-state accuracy.

\begin{figure}[t]
    \centering
    \begin{subfigure}{0.45\linewidth}
      \centering
      \includegraphics[width=\linewidth]{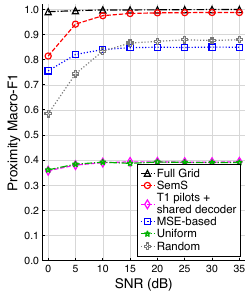}
      \caption{Macro-F1 for proximity recovery.}
      \label{fig:proximity_recovery}
    \end{subfigure}
    \hfill
    \begin{subfigure}{0.45\linewidth}
      \centering
      \includegraphics[width=\linewidth]{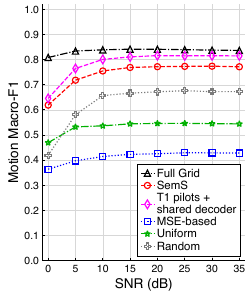}
      \caption{Macro-F1 for radial-motion recovery.}
      \label{fig:motion_recovery}
    \end{subfigure}
    \caption{Performance comparison for semantic-attribute recovery.}
    \label{fig:attribute_recovery}
\end{figure}

Fig.~\ref{fig:attribute_recovery} further compares the recovery of proximity and radial motion from the shared semantic representation. As shown in Fig.~\ref{fig:attribute_recovery}(\subref{fig:proximity_recovery}), SemS achieves higher proximity Macro-F1 than the pilot-limited benchmarks and approaches the Full Grid reference as SNR increases. MSE-based and Random also benefit from increasing SNR, while Uniform shows little improvement. In Fig.~\ref{fig:attribute_recovery}(\subref{fig:motion_recovery}), the gap between SemS and MSE-based becomes more pronounced for radial-motion recovery. This is because SemS optimizes pilot placement directly for semantic-attribute recovery, while MSE-based prioritizes reconstruction of the complete channel response. MSE-based recovers proximity relatively well but yields the lowest radial-motion Macro-F1. Since approach warning depends on both attributes, its weaker radial-motion recovery also helps explain its lower warning performance. The shared semantic objective instead supports both task outputs through recovery of their common representation.\par

\subsection{Pilot Budget and RF Impairments}
\label{sec_result_budget_rf}
{To investigate the effect of sensing resources, we evaluate the impact of the pilot budget $N_p$ on the performance of the two tasks.} For each budget, the same SemS pilot pattern and semantic decoder support both target classification and approach warning. SemS is compared with MSE-based, Uniform, and Random under matched pilot counts.
\begin{figure}[tb]
    \centering
    \begin{subfigure}{0.45\linewidth}
      \centering
      \includegraphics[width=\linewidth]{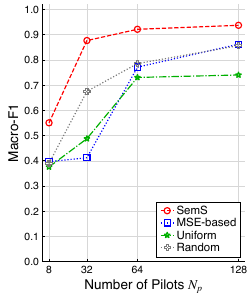}
      \caption{Target classification.}
      \label{fig:pilot_budget_classification}
    \end{subfigure}
    \hfill
    \begin{subfigure}{0.45\linewidth}
      \centering
      \includegraphics[width=\linewidth]{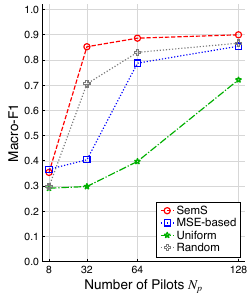}
      \caption{Approach warning.}
      \label{fig:pilot_budget_warning}
    \end{subfigure}
    \caption{Impact of pilot budget $N_p$ on task Macro-F1 at 20 dB.}
    \label{fig:pilot_budget_impact}
\end{figure}
Fig.~\ref{fig:pilot_budget_impact} illustrates how the pilot budget affects target classification and approach warning. As shown in Fig.~\ref{fig:pilot_budget_impact}(\subref{fig:pilot_budget_classification}), SemS achieves higher classification Macro-F1 than MSE-based, Uniform, and Random at the displayed budgets. MSE-based and Random continue to improve as more pilots become available, while the classification Macro-F1 of Uniform levels off toward the upper end of the displayed range. Fig.~\ref{fig:pilot_budget_impact}(\subref{fig:pilot_budget_warning}) shows that SemS and MSE-based achieve similar approach-warning performance at the smallest pilot budget, while SemS retains its advantage at the larger pilot budgets. For Uniform, increasing the pilot budget at the upper end brings a larger gain in warning performance than in classification. The improved proximity recovery is consistent with this gain, as proximity contributes to warning decisions while target classification depends on object category. The SemS curves level off at larger pilot budgets in both tasks, indicating diminishing gains from further increasing the number of pilots. Future designs could dynamically allocate sensing resources and select inference modes to balance semantic accuracy, sensing overhead, and energy consumption \cite{liu2024mode}. Online adjustment based on received observations could further adapt these decisions to changing scenes and task requirements.

\begin{figure}[t]
    \centering
    \begin{subfigure}[t]{0.45\linewidth}
      \centering
      \includegraphics[width=\linewidth]{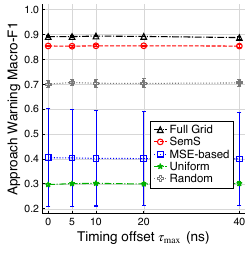}
      \caption{Residual timing offsets.}
      \label{fig:residual_timing}
    \end{subfigure}
    \hfill
    \begin{subfigure}[t]{0.45\linewidth}
      \centering
      \includegraphics[width=\linewidth]{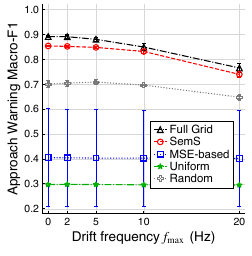}
      \caption{Slow-time phase drift.}
      \label{fig:slow_time_drift}
    \end{subfigure}
    \caption{Sensitivity of approach-warning Macro-F1 to RF impairments at 20 dB. Curves and error bars show means and sample standard deviations over three model seeds, respectively.}
    \label{fig:residual_phase_sensitivity}
\end{figure}

% BEGIN Andrew R5.6 approved 2026-09-22
{
We examine approach-warning sensitivity to RF impairments with the trained models unchanged. Residual timing offsets and slow-time phase drift are tested separately using framewise offsets or drift frequencies drawn uniformly from symmetric intervals with varying maximum magnitudes. Fig.~\ref{fig:residual_phase_sensitivity}(a) shows little change for maximum timing offsets up to 40 ns, with SemS maintaining a Macro-F1 near 0.85. SemS outperforms the other pilot-limited methods, while Full Grid retains a higher score. A framewise timing offset preserves the slow-time phase progression used for radial-motion recovery. In Fig.~\ref{fig:residual_phase_sensitivity}(b), phase drift reduces the Macro-F1 of SemS, Random, and Full Grid, with SemS decreasing from 0.853 to 0.739 at a maximum drift frequency of 20 Hz while remaining above the other pilot-limited methods. Phase drift perturbs Doppler-related cues, and the decline of Full Grid indicates that dense observations alone do not eliminate this sensitivity. These results motivate residual-phase estimation and compensation or training with sampled phase errors.

\par}

% END Andrew R5.6 approved 2026-09-22

\subsection{SemS-based Range Estimation}
\label{sec_result_C}
The ranging experiments use the target model in Table~\ref{tab:range_velocity_micro} and the OFDM parameters specified above. Range mean absolute error (MAE) is averaged over three runs on 3,240 independently generated test frames. Uniform pilots, Random, and ambiguity-function-optimized (AF-optimized) pilots use the same receiver architecture as SemS, with parameters trained separately for each pilot pattern. Uniform pilots and Random sample subcarriers at evenly spaced time slots, with Random retaining both band edges. AF-optimized pilots suppress delay-ambiguity sidelobes. We also compare with orthogonal matching pursuit (OMP) and multiple signal classification (MUSIC) \cite{andrew2021overview}. OMP and MUSIC-OP use the AF-optimized pattern, MUSIC-UP uses uniform sampling, and MUSIC-RP randomly selects subcarrier and time indices. With clutter, these traditional methods use temporal differencing before delay estimation.

\begin{figure}[tb]
    \centering
    \begin{subfigure}{0.45\linewidth}
      \centering
      \includegraphics[width=\linewidth]{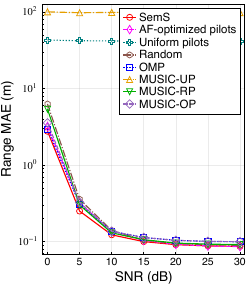}
      \caption{Without clutter.}
      \label{fig:ranging_mae_without_clutter}
    \end{subfigure}
    \hfill
    \begin{subfigure}{0.45\linewidth}
      \centering
      \includegraphics[width=\linewidth]{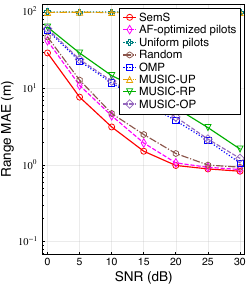}
      \caption{With clutter.}
      \label{fig:ranging_mae_with_clutter}
    \end{subfigure}
    \caption{Range MAE comparison with $N_p=32$.}
    \label{fig:ranging_mae}
\end{figure}

\begin{figure}[tb]
    \centering
    \includegraphics[width=0.55\linewidth]{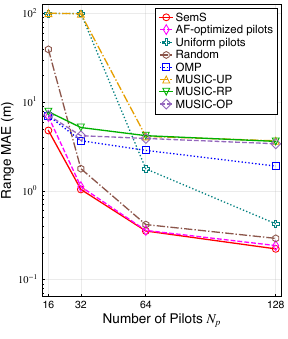}
    \caption{Range MAE versus the number of pilots with clutter at 20 dB.}
    \label{fig:ranging_pilot_budget}
\end{figure}

Fig.~\ref{fig:ranging_mae} compares the range MAE across varying SNRs. In Fig.~\ref{fig:ranging_mae}(\subref{fig:ranging_mae_without_clutter}), the errors of SemS, AF-optimized pilots, Random, OMP, MUSIC-RP, and MUSIC-OP decrease rapidly as SNR increases, followed by smaller improvements at high SNR. SemS achieves lower MAE than OMP and the MUSIC variants, while its high-SNR performance is close to that of AF-optimized pilots and Random. Uniform pilots and MUSIC-UP retain large errors with little improvement. Their sparse, regularly spaced frequency samples produce repeated delay responses within the delay search interval, leaving a range ambiguity that increasing SNR alone cannot resolve. In the cluttered case shown in Fig.~\ref{fig:ranging_mae}(\subref{fig:ranging_mae_with_clutter}), SemS achieves lower MAE than all the compared baselines throughout the tested SNR range. AF-optimized pilots and Random also benefit from increasing SNR, whereas OMP and the MUSIC variants exhibit larger errors.

Fig.~\ref{fig:ranging_pilot_budget} further compares the ranging errors under different pilot budgets in clutter at 20 dB. SemS achieves the lowest mean MAE at all displayed budgets. At $N_p=16$, its MAE is 30.4\% lower than that of AF-optimized pilots, while the two methods are close at $N_p=64$. At each budget, all methods use the same energy per pilot. AF-optimized pilots suppress delay-ambiguity sidelobes, whereas SemS jointly adjusts pilot placement and the receiver through the range-estimation loss. This task-driven design directs the available pilot energy toward acquiring task-relevant information and adapts the receiver to use these observations in clutter. The comparison supports joint pilot and receiver optimization for range estimation with limited pilot resources.

\section{Conclusion}
We proposed a SemS framework that extends task-oriented sensing by explicitly modeling and recovering the environmental information required for downstream inference. The information-theoretic criterion guides the joint design of transmit waveforms and receiver processing toward recovery of task-relevant information. For a given probing waveform, we characterize semantic recovery by averaging the decoder's output distribution over physical realizations sharing the same semantic state. We then developed a DL-based OFDM implementation that jointly optimizes a pilot pattern and semantic decoder to recover a shared representation, which supports target classification and approach warning through predefined mappings without task-specific retraining. Numerical results demonstrate that SemS achieves higher target-classification accuracy and approach-warning Macro-F1 than reconstruction-based pilot design under the same pilot budget. Future research will extend this framework to unified semantic ISAC systems and use radar measurements to evaluate its performance across broader target and channel conditions and under practical RF impairments.

\bibliographystyle{IEEEtran}
\bibliography{ref}

\end{document}